\documentclass{article}

\usepackage{amsmath}

\usepackage{mathptmx}           
\usepackage{graphicx}           
\usepackage{url}                

\usepackage[a4paper,margin=2cm]{geometry}

\usepackage{natbib}
\bibpunct{(}{)}{;}{a}{,}{,}

\usepackage{textcomp}

\usepackage{xspace}

\begin{document}

\author{ Nico Reinke$^{\small{1,2,4}}$
  \\
  \\
$^{\small{1}}$ Institute of Physics,  University of Oldenburg and
\\
$^{\small{2}}$ ForWind,  University of Oldenburg, Carl-von-Ossietzky-Str. 9-11, 26129 Oldenburg, Germany
\\
$^{\small{4}}$ Corresponding author - Email: nicoreinke@gmx.de
  }
\title{A spatially optimized wind tunnel}
\maketitle

\begin{abstract}
In this report a wind tunnel is described in term of its design, its construction as well as in its validation of flow features. 
This wind tunnel is characterized by a new unconventional design.
The innovative design allows an unusual long flow test section (in particular important for turbulence research) relative to the total size of the wind tunnel.
Thus, the new design enables a much better use of space in comparison to a wind tunnel with conventional design.
The design is characterized by two strong expansions of flow, namely a  {wide angle diffusor}  and  strong expanding corner.
Among other things, investigations about these flow expanding components enlarge the existing knowledge and show new possibilities.
\end{abstract}

\section{Introduction}
\begin{figure}[h]
\centering
\includegraphics[width=1\textwidth]{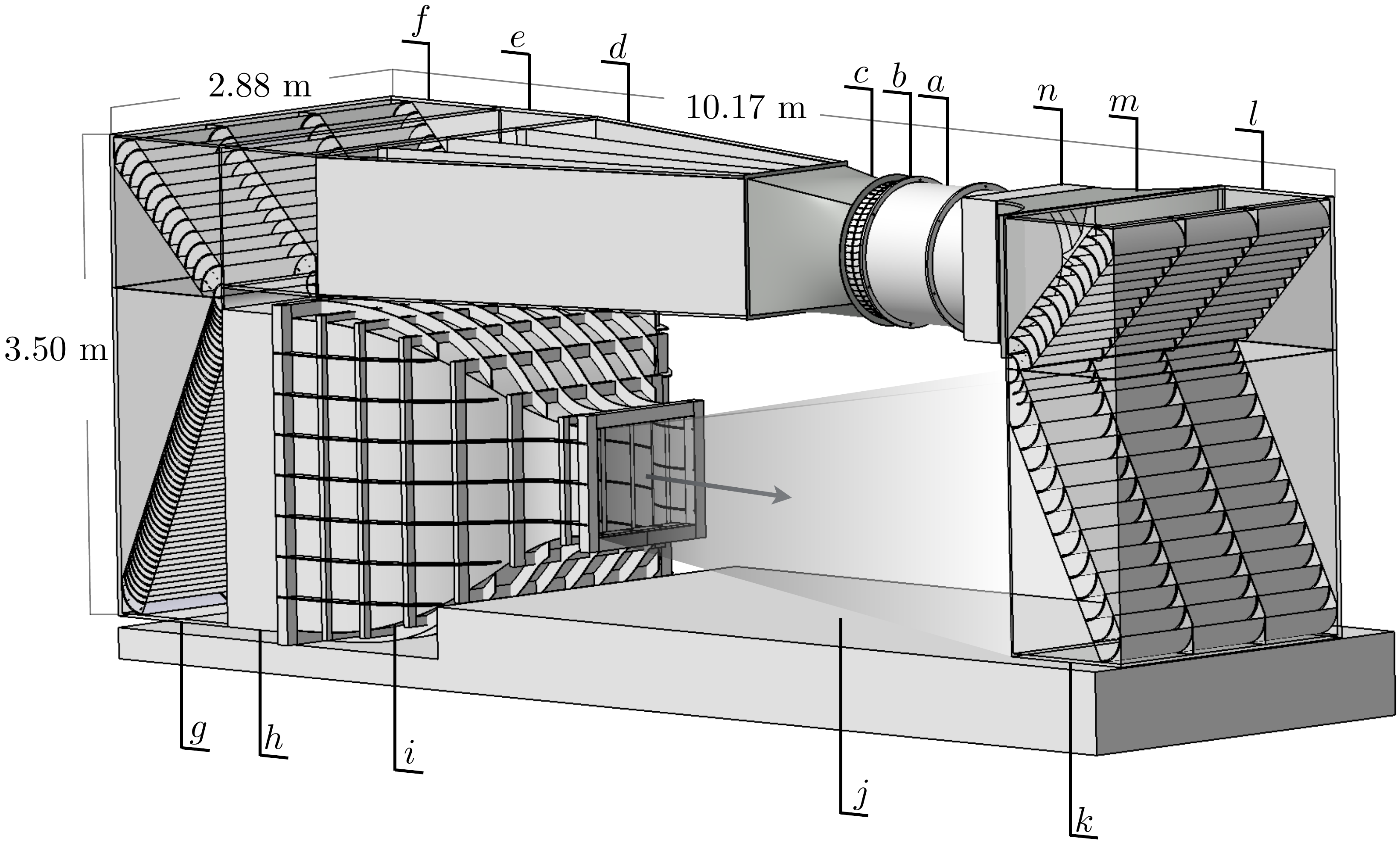}
\vspace{0.5cm}
\caption{
Technical sketch of the wind tunnel. Walls are partly transparent to reveal the inside of tunnel and its flow modulators. Arrow indicates flow direction
}
  \label{fig:wk}
\end{figure}

A G\"ottingen type wind tunnel is designed and built up for turbulent flow research and 
the characterization of objects in such flows.
Due to spatial limitation ($10.5\times3.8\times5$~m$^3$) and the need of a maximal long test section an unconventional wind tunnel design was developed. 
This design enables a test section with dimensions of $5.2\times 1.0 \times 0.8$~m$^3$ (length$\times$width$\times$height).
The  G\"ottingen type or recirculation wind tunnel  is selected since it is characterized by a high efficiency (only the inner friction slows down the flow and has to be overcome by the ventilator power) and its encapsulate flow does not disturb surrounding work in the laboratory.
These two aspects are different to an Eifel wind tunnel type\footnote
{
An Eifel type or linear wind tunnel is not a closed loop tunnel, it  sucks in and blows out the air from the laboratory or the surrounding. 
However, since G\"ottingen and Eifel design have their pros and contras, the concept of wind tunnel construction has to be evaluated on the basis of requirements
}.  
\\
The wind tunnel design enables to work in a closed test section setup as well as in an open setup.
Wind speed reaches up to  \mbox{$u_\infty$~=~22~$\frac{\rm{m}}{\rm{s}}$} in the test section. 
Although an unconventional design is present, the turbulence intensity\footnote
{
Turbulence intensity is determined by the ratio between standard deviation of velocity and temporal mean velocity, $TI=\frac{\sigma(u)}{\langle u \rangle_t}$ 
}
 is within a range of 0.7\%~$\le~TI~\le$~2.9\% for various velocities and along the centerline of the closed test section.
\\
This wind tunnel design is possible due to a {wide angle diffuser} and a {strong expanding corner}.  
The wide angle diffuser is constructed with four sub-diffusers, which prevent flow separation during flow expansion and keep the level of turbulence intensity low.
Moreover, the concrete split of the flow within the diffuser allows adjusting the velocity field in terms of its profile. 
\\
Commonly, in G\"ottingen type wind tunnels flow is guided around corners with so-called {guide plates}.
Unfortunately, literature states only a rule of thump for the installation of guide plates and in particular, expanding corners are  discussed very little.
Therefore, flow features of guide plates are studied experimentally (with and without expansion).
The wake of guide plates is analyzed in terms of mean velocity and its turbulence intensity.
For the first time, results show how wake features are  dependent on inflow conditions (Reynolds number)
as well as on distance of adjacent guide plates.
In particular, results present what wake features can be expected.
This new knowledge can be used during the design procedure of a wind tunnel, which is also done in this chapter.
\\
A technical drawing of the wind tunnel is shown in figure  \ref{fig:wk}.
Note, distance between wind tunnel and laboratory walls or other limitations are in a range of 10~cm~-~15~cm. 
The air flow\footnote{
Air in  ambient conditions ($T~\approx$~20$^\circ$C, $p~\approx$~1000~hPa) is the flow media of the present  wind tunnel}
streams through the tunnel from component \textit{a, b, c ...} to \textit{n}.
The cycle of air starts at the ventilator (component \textit{a}), which accelerates the flow.
Immediately thereafter, the air streams through a flow straightener (component \textit{b}).
Then the wind tunnel cross section changes from circular to square (component \textit{c}) and the air stream gets expanded in the wide angle diffuser (component \textit{d}).
Behind the diffuser, honey combs are installed in the tunnel, these damp cross flow fluctuation (component \textit{e}).
Afterwards, the flow is guided around two $90^\circ$ corners (component \textit{f} and \textit{g}).
Component \textit{g} is the strong expanding corner and then flow arrives the settling chamber (component \textit{h}). In the settling chamber the flow streams through three flow modulators, namely two grids and one net. Thereby, flow fluctuations and velocity gradients get damped.
Thereafter, flow is accelerated in a nozzle (component \textit{i}). 
Then the flow reaches the open test section (component \textit{j}). 
The open test section can be converted/rebuilt to a close test section. 
At the end of the test section, flow gets collected and guided around two $90^\circ$ corners  (component \textit{k} and \textit{l}).
Then flow gets contracted in a simple pre-nozzle (component \textit{m}) and in a transition, where the wind tunnel cross section morphs from square to circular (component \textit{n}) and the flow comes back to the ventilator. 
\\
Below, the single components  of the wind tunnel (\textit{a} - \textit{n}) are described. 
Thereby, their construction, their adjustments and corresponding experiments are explained and discussed.
Afterwards, flow quality of the test section is shown in terms of velocity features and turbulence intensity. 
Finally, results as well as optimization possibilities of the presented wind tunnel are summarized and discussed.

\section{Wind tunnel components}
\label{sec:components}
\subsection*{a: Ventilator}
A ventilator accelerates the flow in the wind tunnel, component \textit{a} in figure \ref{fig:wk}.
This ventilator is an axial type, which is characterized by a high energy efficiency, therefore it is commonly used in wind tunnels, cf.  \cite{Goodrich}.
The inner diameter of the ventilator is 1.01~m. 
The maximal volume flux is $\dot Q$~=~16.6~$\frac{\rm{m^3}}{\rm{s}}$, which corresponds to a test section inflow velocity of \mbox{$u_\infty~\approx$~20.8~$\frac{\rm{m}}{\rm{s}}$.} 
This is an idealized manufacturer specification without any additional pressure drop\footnote
{
The pressure drop $\Delta p$ is defined by the difference of pressure measured at two spatial separated points in a flow. It is caused by frictional forces and indicates the loss of potential energy of the flow while passing of body or a tube, cf. \cite{Schlichting2000}  
}. 
The maximal power consumption is 15~kW, though, different settings enable a  maximal power consumption of 17~kW, which is not done in the current setup.
Another characteristic of a ventilator is the so-called {pressure resistance}. 
The pressure resistance characterizes volume flux features of the ventilator to a certain pressure drop. 
In case the pressure drop of the wind tunnel changes also the volume flux changes. 
It changes less when the pressure resistance of the ventilator is high. 
In particular, for turbulence experiments with an active grid a ventilator with high pressure resistance is beneficial, since an active grid in operation has commonly a time dependent pressure drop due to its dynamically change of blockage. 
For a reliable and stable flow modulation with an active grid, without any hysteresis-effects or wind tunnel pumping, a ventilator with high pressure resistance is selected.
It is build up with 12 broad blades. 
Note, the  pressure resistance results from the number of blade as well as its broadness.
\\
This ventilator is selected because of its maximal volume flux and the strong pressure resistance. 
The wake velocity profile of the ventilator was less considered in the choice of the ventilator. 
Its wake velocity profile is not uniform and contains a strong vortex. The equalization of this issue gets further discussed below.

\subsection*{b: Vortex straightener}
\begin{figure}[h]
\centering
\includegraphics[width=0.8\textwidth]{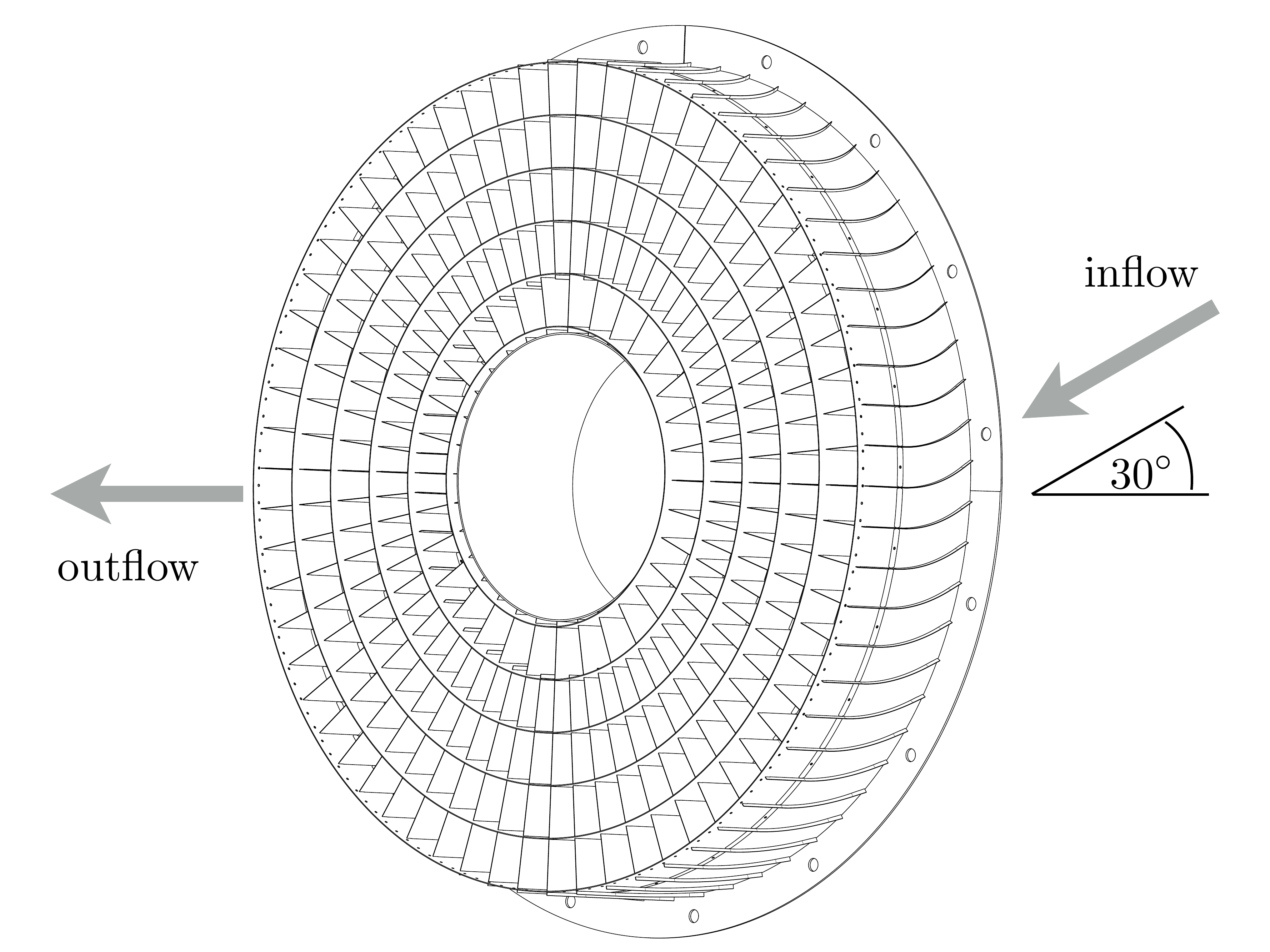}
\vspace{0.5cm} \caption{Technical drawing of the vortex straightener. The component downstream length is 20~cm. Due to the hub of the ventilator and its wake no plates are installed in the center of straightener}
  \label{fig:Vortex_destroyer}
\end{figure}

The non-uniform and rotating wake of the ventilator gets improved by a flow straightener, figure \ref{fig:Vortex_destroyer}.
This flow straightener is built up to equalize the flow vortex.
The flow straightener is located right behind the ventilator, component \textit{b} in figure \ref{fig:wk}.
\\
Measurements have shown that the flow comes with an inclination angle of approximately 30$^\circ$ out of the ventilator. 
The flow straightener is build up in such a way that the flow is collected by this inclination angle and inside, the flow is guided to  inclination angle of 0$^\circ$, 
thus the flow becomes parallel to the wind tunnel centerline. 
Due to this specific collection of the flow, 
flow separation is prevented at the guide plates\footnote{
Complete flow separation is prevented on a straight plate up to an angle of attack of approximately $10^\circ$, cf. \cite{Mehta}}.
\\
Wind tunnel performance is tested in dependency from the number of guide plates in the flow straightener. 
A configuration with 64 guide plates shows the best flow result. 
(Flow features are measured behind the diffuser (component \textit{d}).)
For instance, 32 guide plates result in a higher turbulence intensity and do not increase the volume flux.
Although 32 guide plates should induce a lower pressure drop than the configuration with 64 guide plates.
One might assume that the diffuser works less efficient, since the flow is more disturbed with less guide plates.
\\
Rings are integrated in the straightener construction, which enhance stability of the straightener. 
Moreover, these rings reduces radial flows. 
The reduction of these radial flows improves the equalization of the non-uniform velocity profile on its way downstream.
The importance of the flow straightener in terms of flow features becomes obvious in section d: Wide angle diffuser.

\subsection*{c: Transition from circular to squared cross section}
The circular cross section of the vortex straightener morphs to the squared cross section of the diffuser in component \textit{c} figure \ref{fig:wk}.
The transition takes place in the four corners of the component. 
Each corner corresponds to a quarter circle of the incoming cross section.
In a linearly fashion this circular shape morphs to right-angle corner.
This morphing is equivalent to a flow expansion of approximately 1:1.27 (ventilator diameter 1.01~m, inlet edge length of diffuser 1.01~m).
Geometrical features of the transition are designed as suggested in \cite[p. 7]{Kuehle} and \cite{Kline} in order to avoid flow separation.
Thereby, the opening angle of the transition as well as the ratio between length of the transition and its broadness determine whether or not flow separation takes place.   
The mentioned ratio is approximately one in component \textit{c}, which leads to a maximal opening angle of approximately $\pm$~16$^\circ$ where no flow separation is expected.
According to this recommendation, the maximal opening angle of the transition is set to $\pm$~11.8$^\circ$.

\subsection*{d: Wide angle diffuser}
\begin{figure}[h]
\centering  
a)\includegraphics[width=0.45\textwidth]{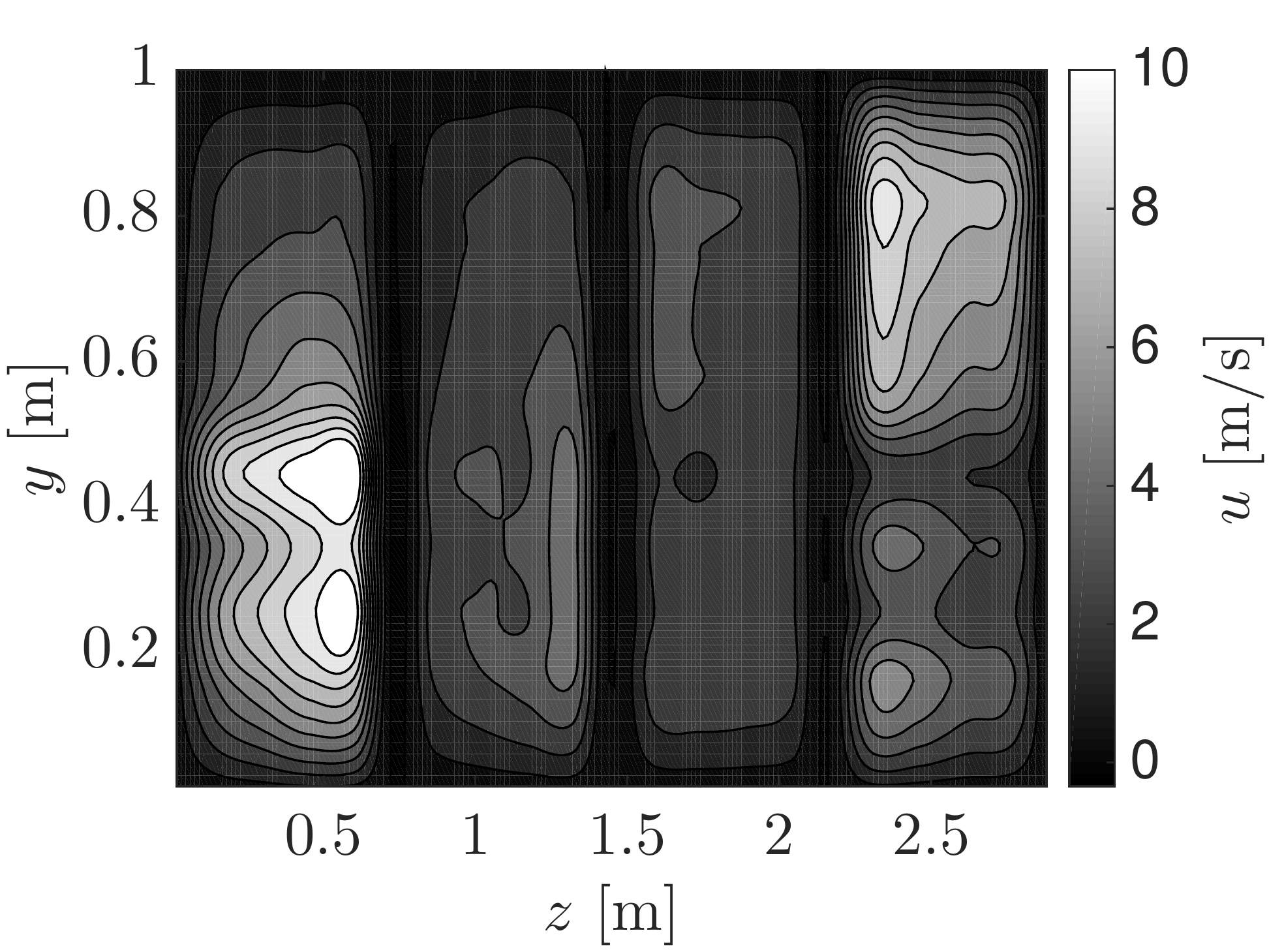}
b)\includegraphics[width=0.45\textwidth]{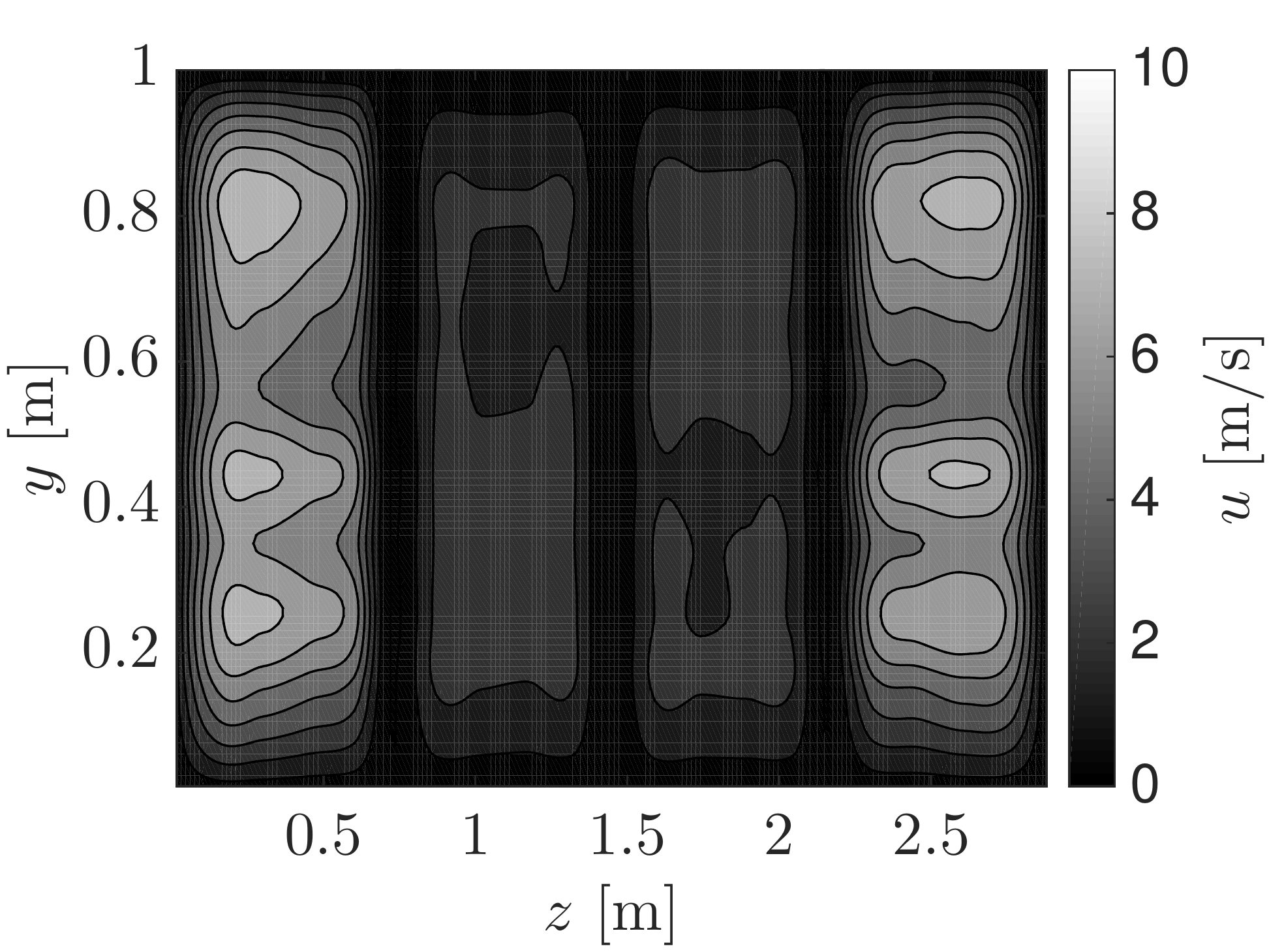}
c)\includegraphics[width=0.45\textwidth]{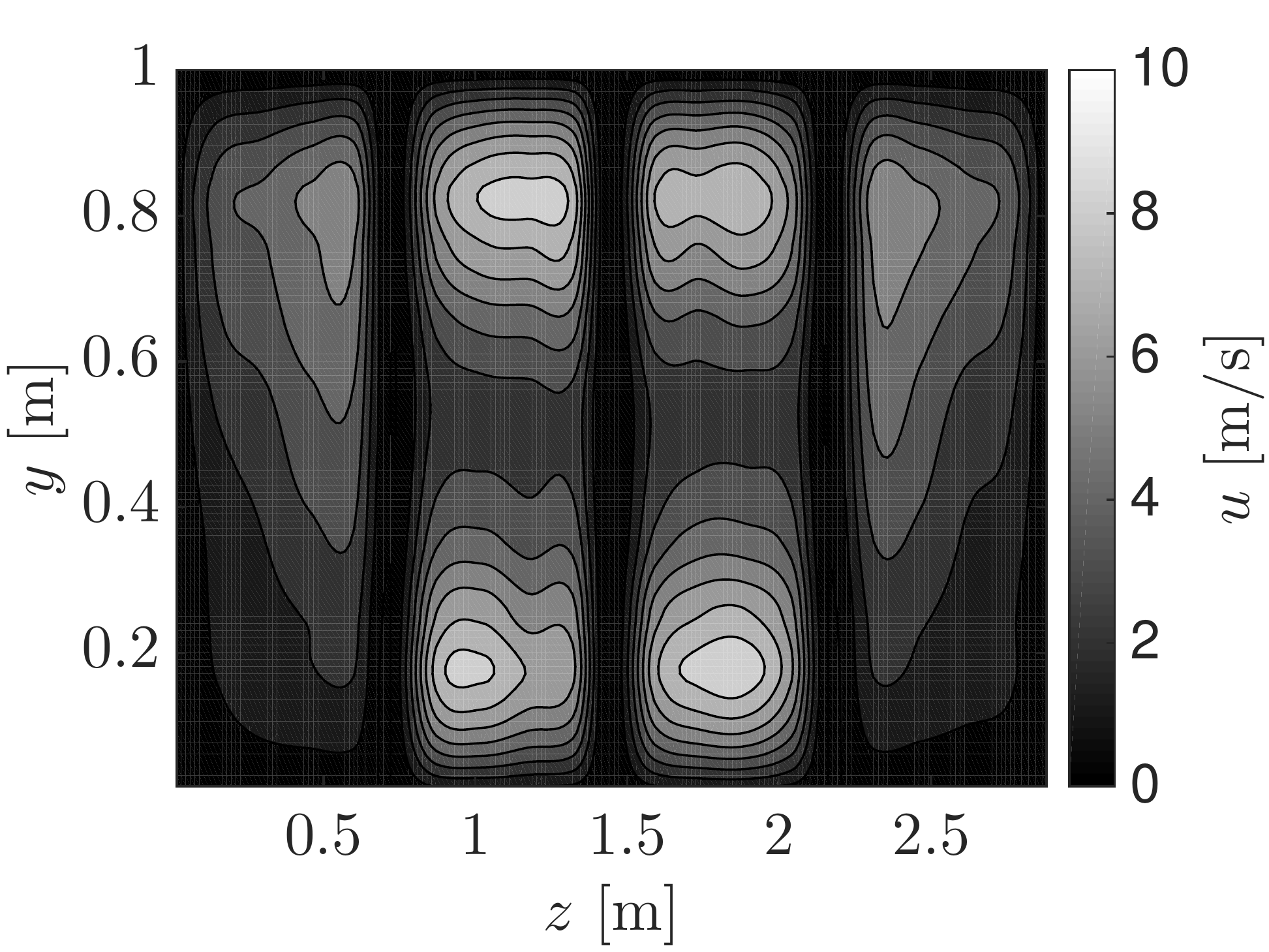}
d)\includegraphics[width=0.45\textwidth]{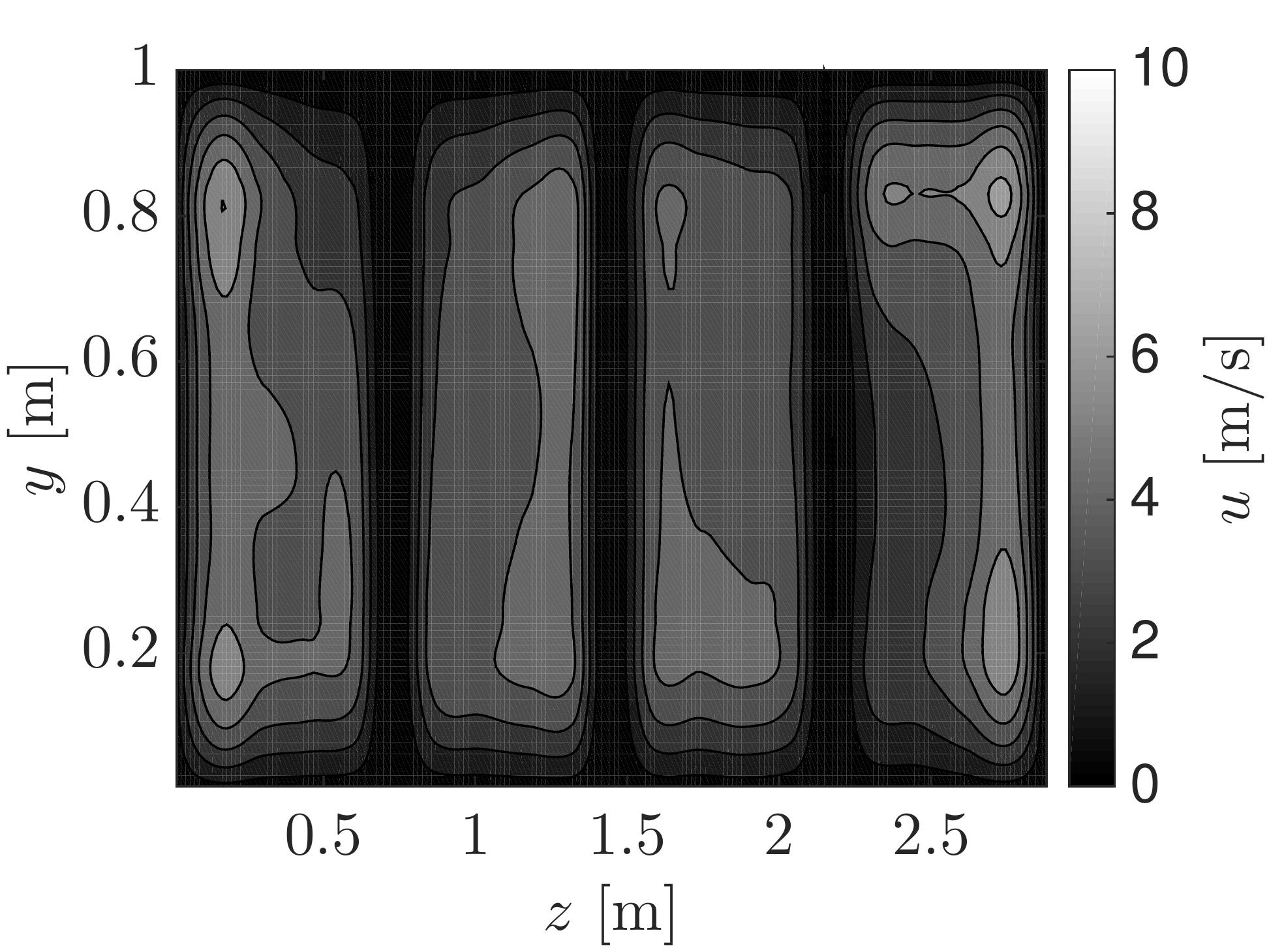}
\vspace{0.5cm} \caption{
Velocity fields behind the wide angle diffuser.
a) unadjusted diffuser, without vortex straightener, 
b) wake with vortex straightener, 
c) first adjustment of volume flux,
d) final adjustment with wake cone}
\label{fig:diffuser}
\end{figure}

The wide angle diffuser is component \textit{d} in figure \ref{fig:wk}.
On the one hand, it is of particular importance to avoid flow separation within the diffuser, since it already influences test section flow features as well as efficiency of the wind tunnel.
Therefore, \cite[p. 7]{Kuehle} and \cite{Kline} suggest design rules to avoid flow separation.
Two crucial quantities are worked out for the design of linear diffuser, the opening angle of the diffuser and the ratio of the length of the diffuser to its inlet width.   
On the other hand,
the wind tunnel requirements and in particular the pre-defined dimensions of the nozzle (component \textit{i}) lead to fixed diffuser dimensions of 3.50~m length, inlet width \mbox{1.01~m},  outlet width 2.88~m as well as a constant height of 1.01~m.
The resulting opening angle is approximately $\pm$~14.3$^\circ$, which indicates flow separation for this length to inlet width ratio of roughly 3.47, cf. \cite[p. 7]{Kuehle} and \cite{Kline}.
To decrease the opening angle three walls are installed 
in the diffuser, which divide the wide angle diffuser in four sub-diffusers. 
The walls equally divide the inlet and outlet width of the diffuser, thus, the flow gets expanded by an opening angle of $\pm$~3.6$^\circ$ in the sub-diffusers. 
The geometrical features of the sub-diffusers match the suggestions made by \cite[p. 7]{Kuehle} and \cite{Kline}.
Similar approaches to avoid flow separation are mentioned in literature, e.g. \cite{Mehta,Cheng}.
\\
The four sub-diffusers are adjusted in such a way that the velocity profile behind the diffuser is as uniform as possible. 
As mentioned, the velocity profile at this location is already important for the test section flow features.
The four sub-diffusers are iteratively adjusted on the basis of single hot-wire measurements behind the diffuser. 
Measurements were conducted at $5\times8$ (width $\times$ height) evenly distributed positions over each sub-diffuser cross section. 
In postprocessing, the 160 measurements are interpolated along the cross section of the diffuser.
\mbox{Figure \ref{fig:diffuser} a) - d)} shows exemplary temporal mean velocity measurement results, which are conducted during the iterative adjustment process in the diffuser wake. 
Figure \ref{fig:diffuser} a) shows the velocity profile without the vortex straightener (component \textit{b}).
Thereby, two unfavorable high speed regimes exist in the two outer sub-diffusers. 
Most likely, this  high speed regimes are related to the ventilator induced vortex and the non-uniform velocity profile also induced by the ventilator.
This interpretation leads to the construction and installation of the vortex straightener.
Figure \ref{fig:diffuser} b) presents the velocity profile with integrated vortex straightener. 
Thereby, high speed regimes expand over the whole cross section of the two outer sub-diffusers, which might be related to the reduced rotation of the flow behind the vortex straightener.
Afterwards, sub-diffuser entrances are adjusted on the basis of the ratio between volume flux and entrance dimensions of prior result, figure \ref{fig:diffuser} b).
This adjustment is realized by {flexible} PVC-plates at the beginning of sub-diffuser walls, these extensions of walls are located in the transition (component \textit{c}).
The resulting velocity profile is shown in figure \ref{fig:diffuser} c).
Due to the wake deficit in the center of the velocity profile, it becomes evident that the wake of the ventilator hub still influences the diffuser wake.
Therefore, a wake cone is installed behind the ventilator hub, with a length of approximately 1~m.
In addition to the wake cone, new plates adjustments are done before the final equalized velocity profile is reached.
Figure \mbox{\ref{fig:diffuser} d)} shows this velocity profile. 
For such a {strong} flow expansion and such a {short} downstream positions (distance to the ventilator 4.8~m) the degree of flow equalization is already very good.
Although, velocity profile is quite uniform and geometrical suggests are applied, turbulence intensity is around $TI~\approx$~40\%\footnote
{
At {high} turbulence intensities the intrinsic error of the hot-wire measurement technic is increased. 
Therefore, $TI~\approx$~40\% is more a rough estimate than an exact value. 
Due to the fact that hot-wire probes only measure absolute values of the velocity, negative velocities are not recorded and registered as positive ones. 
For instance, that leads to a decreased signals standard deviation ($\sigma(q)=\sqrt{\langle q - \langle q \rangle \rangle^2}$) in comparison to the real velocity field and inversely for the mean velocity. 
Thus, turbulence intensity gets underestimated
}. 
This high level of turbulence intensity can be reduced by grids, screens or honey combs, but also by a more filigree splitting of the diffuser, in other words, additional walls inside the sub-diffusers, see also below. 
Note, measurements are also performed for different velocities, these results qualitatively agree with the presented results in \mbox{figure \ref{fig:diffuser}.}

\subsection*{e: Chamber of flow straightener}
Due to the strong turbulence intensity in the wake of the diffuser and to the minimization of the settling chamber length, so-called {honey combs} (flow straightener) are installed behind the diffuser, component \textit{e} in figure \ref{fig:wk}.
The name honey comb derives from its shape.
Since flow straightener walls stand in a hexagonal shape to each other.
Flow straighteners with a honey comb structure are the most efficient kind of flow straighteners, cf. \cite{Kulkarni}.
The honey combs reduce lateral velocity fluctuations.
Thus, turbulence intensity of the flows gets reduced and following guide plates get a more equalized flow\footnote
{ 
Below investigations of guide plates are shown. 
These are performed with a quite laminar inflow ($TI~\approx$~2\%), thus the resulting wake of guide plates may change its features with higher $TI$. 
However, this influence is not studied here
}.
\\
The honey combs geometry is chosen according to the suggestions of \cite{Mehta,Bradshaw,Kulkarni}. However, since honey combs are not customized small deviations are accepted from suggestions.
The diameter of a single comb is chosen in relation to the incoming turbulent flow and its integral length\footnote
{
Integral length is determined by integrating autocorrelation function of velocity fluctuations, cf. \cite{Batchelor1953}. Integration is performed from zero correlation length up to the first zero crossing, which is a common procedure
}. 
Measurements in the wake of the diffuser (similar to these shown in figure \ref{fig:diffuser}) are done to find the integral length distribution. 
these measurements show a mean integral length over the diffuser wake around  $\langle L \rangle$ $\approx$ 0.4~m~-~0.6~m.
The diameter of a honey comb should be smaller than $L$, cf. \cite{Mehta}, other references suggest more precisely that this ratio should be around 8~-~10, e.g. \cite[p. 44]{Kulkarni}. 
Due to limited availability a comb diameter is chosen of roughly 3~cm, which is $\sim$13~-~20 smaller than  $L$.
\\
A more important feature is the ratio between the honey comb diameter and its downstream length.
Commonly, 6~-~12 is chosen for this ratio, cf. \cite{Mehta,Bradshaw,Kulkarni}. 
The installed honey combs have a length of 20~cm, which corresponds to a ratio of $\sim$7.

\subsection*{f \& l: 1. and 4. corner of guide plate}
\begin{figure}[h]
\centering  
\includegraphics[width=0.7\textwidth]{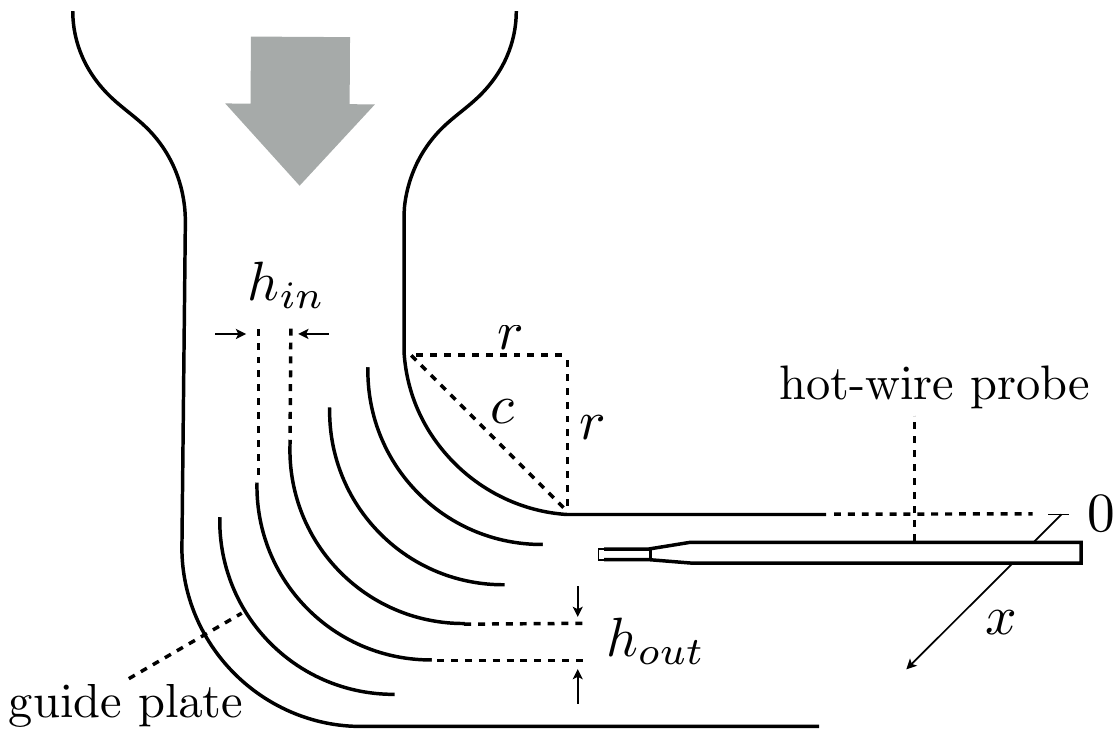}
\vspace{0.5cm} \caption{Experimental setup for investigations on guide plates for $E~=\frac{h_{out}}{h_{in}}=~1$. Massive walls are indicated by solide black lines. Grey arrow shows flow direction. Arrow at the hot-wire probe indicates the displacement x and its \mbox{direction}}
\label{fig:LB_1}
\end{figure}

\begin{figure}[h]
\centering  
a)\includegraphics[width=0.45\textwidth]{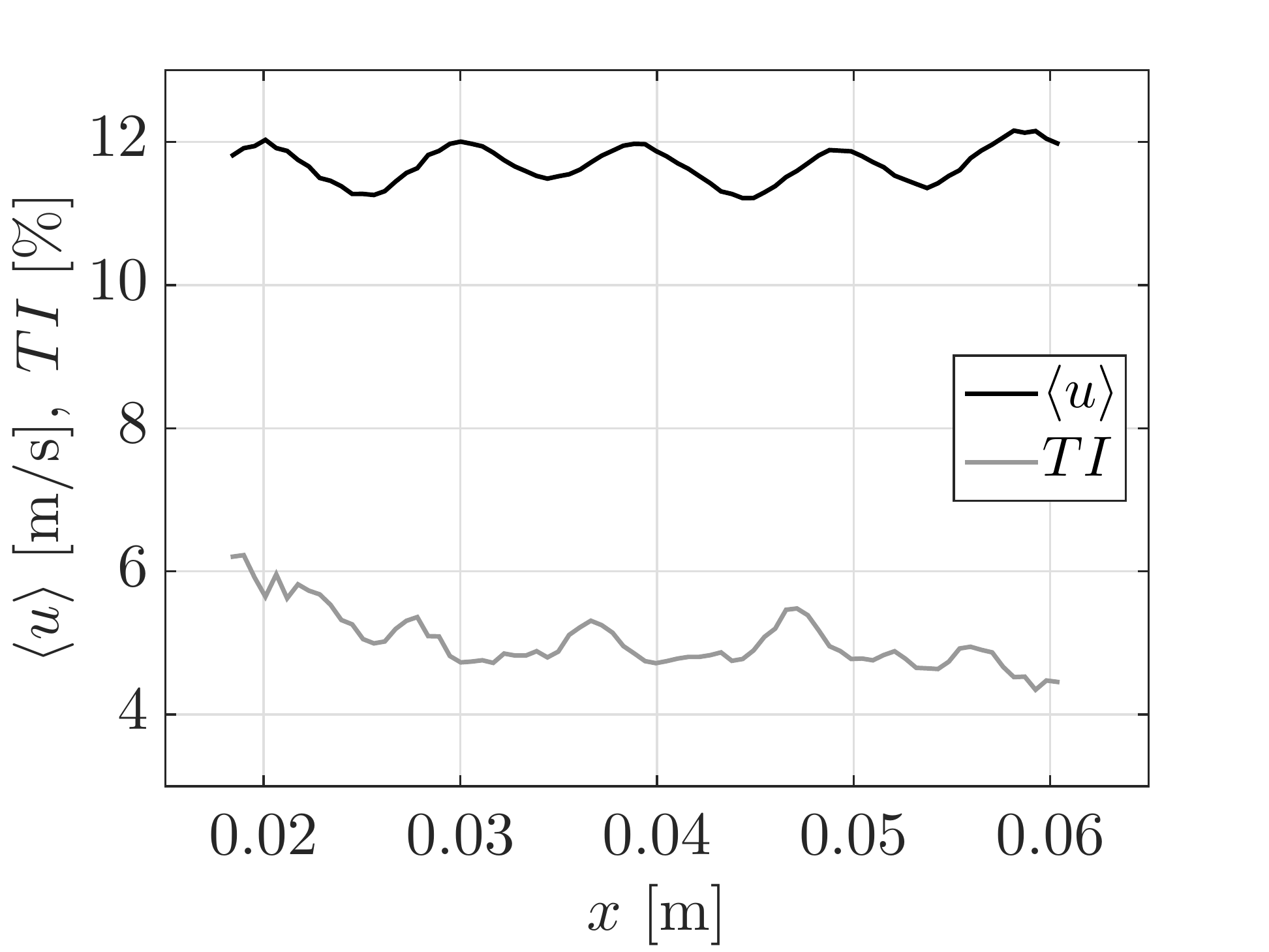}
b)\includegraphics[width=0.45\textwidth]{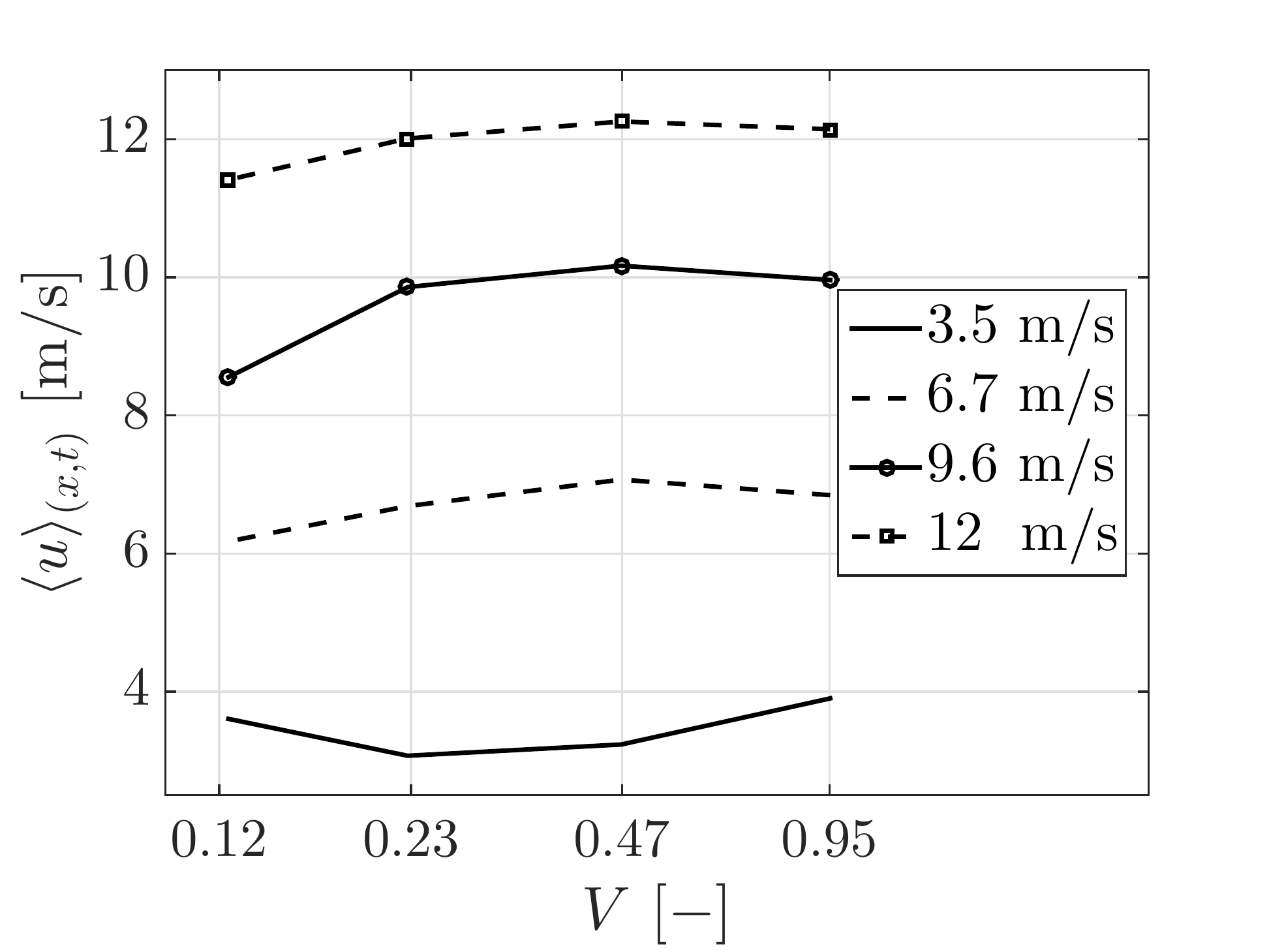}
c)\includegraphics[width=0.45\textwidth]{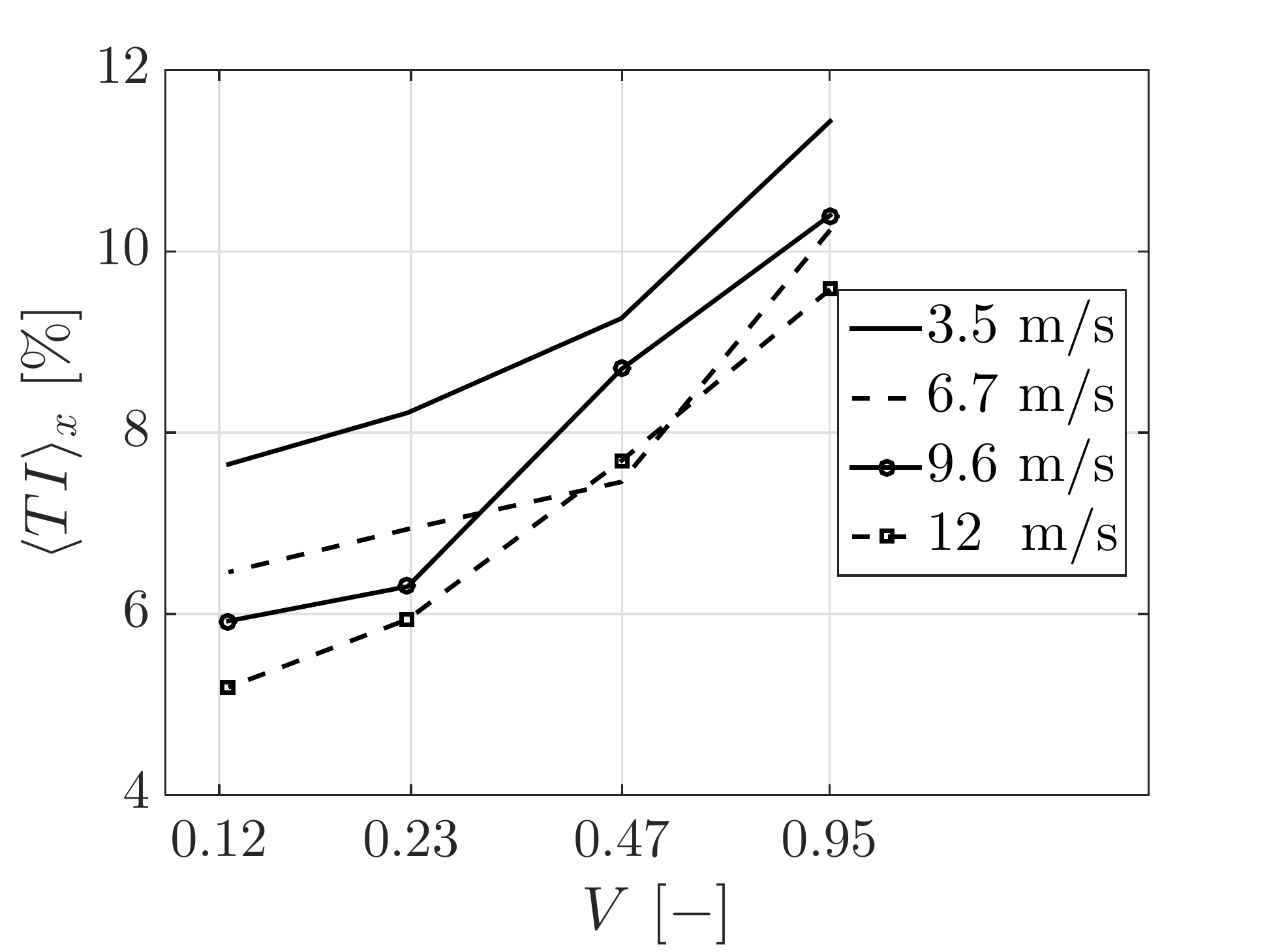}
d)\includegraphics[width=0.45\textwidth]{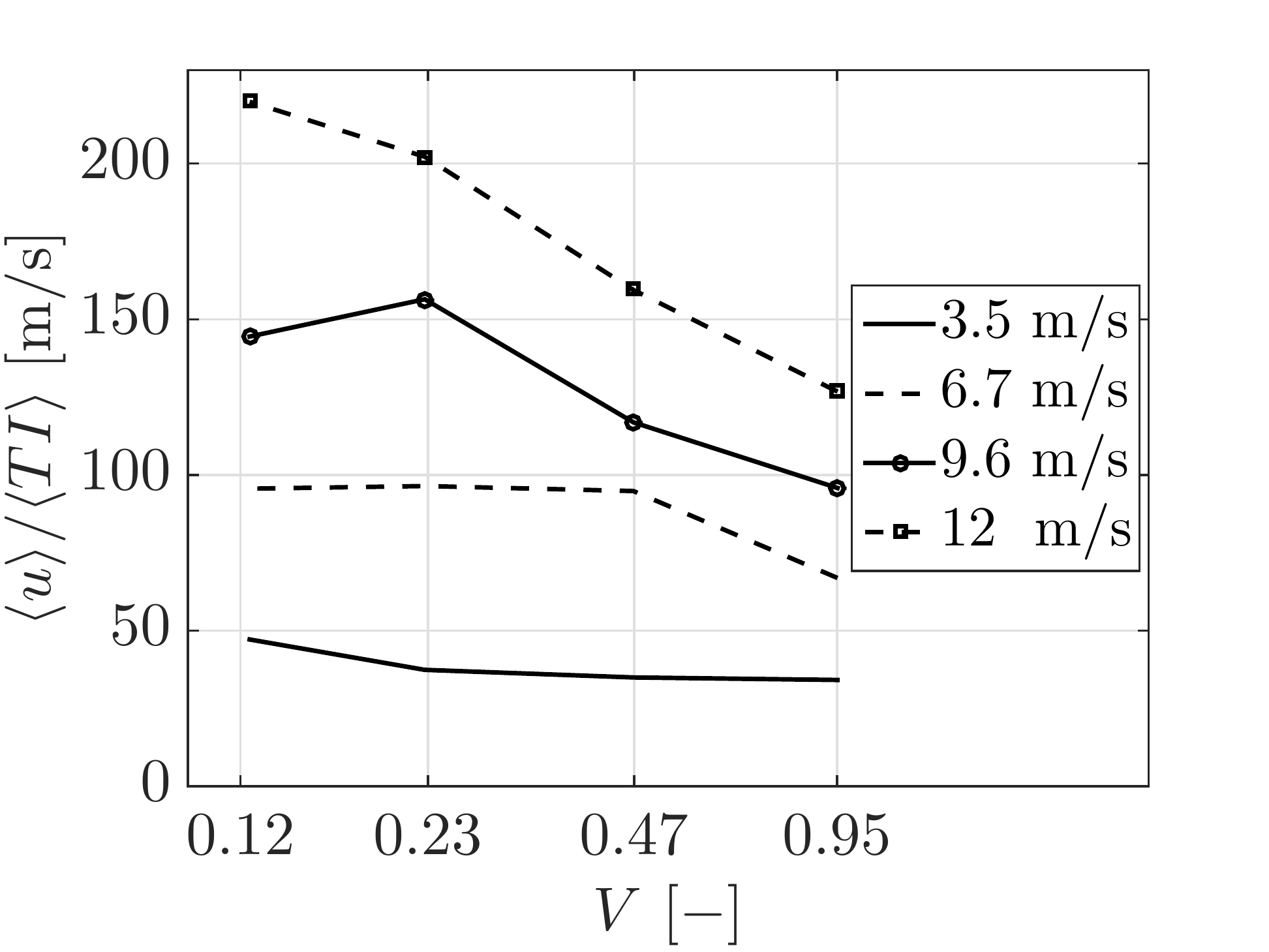}
\vspace{0.5cm} \caption{
Figures summarizes guide plates wake results with $E=1$.
a) temporal mean velocity profile and turbulence intensity profile in the wake of guide plates, $V~\approx~0.12$, 
b) spatial and temporal mean velocity with variation of $V$ and inflow adjustment,
c) spatial mean turbulence intensity with variation of $V$ and inflow adjustment,
d) ratio between spatial and temporal mean velocity b) and spatial mean turbulence intensity c)
}
\label{fig:LB_standard}
\end{figure}

Commonly, four 90$^\circ$ redirect-corners are present in a G\"ottingen type wind tunnel.
For a couple of reasons, it is necessary to prevent flow separation in such a corners, e.g. wind tunnel efficiency, flow features in terms of low turbulence intensity in the test section and noise etc.
For this purpose guide plates are installed in the corners. 
Corners are classified in three groups, 
\begin{itemize}
\item expanding corners: here the flow expands on its way downstream, thus incoming cross section $A_{in}$ is smaller than the going out cross section $A_{out}$, the expansion is $E$~=~$\frac{A_{out}}{A_{in}}~>~$~1;
\item contraction corners, here is $E<1$;
and
\item constant corners, here is $E~=~1$.
\end{itemize}
Flow separation is much more likely in an expanding corner than in a  contracting corner.
Therefore, following investigations are focused on the constant and the expanding case\footnote{
All experiments are performed under environmental or laboratory conditions, e.g. $T~\approx~20^\circ$C; the same working conditions of the here discussed wind tunnel}.
\\
In this section, the constant case is discussed.
In the wind tunnel, guide plates are used with a quarter circular shape and with constant plate thickness of 1~mm.
The only installation recommendation, which is found, is a rule of thump for the installation of such guide plates $V~=~\frac{h}{c}~\leq~0.25$ (for $E~=~1$), cf. \cite{Mehta}. 
Where $h$ is the distance between two plates and $c$ is their chord length.
However, the impact of disregarding this rule of thump as well as 
under which specific conditions 
this rule can be applied is not discussed.  
Thus, experimental investigations are performed to get a more detailed characterization.  
Consequently, the investigations are focused on the ratio $V$ and inflow variations. 
Different inflow velocities correspond to changes of Reynolds number\footnote
{
Reynolds number characterizes the ratio of inertial to viscous forces in the fluid.
With this number flows can be compared in their similarity, although flows are different in dimensions, velocities or viscosities.
It is expected to find similar flow features in {different flows} when their Reynolds number is equal.
Reynolds number is defined by $Re~=~\frac{u d}{\nu}$, where $u$ is a characteristic velocity, $d$ is a characteristic length and $\nu$ is the kinematic viscosity, $\nu~=~\frac{\eta}{\rho}$, where   $\eta$ is the dynamic viscosity and  $\rho$ is the density of flow. 
A $Re$-number for guide plates might be defined based on their chord length $c$, $Re~=~\frac{u c}{\nu}$.
More information on dimensionless Reynolds number can be found in many textbooks, e.g. \cite{Schlichting2000}
}
 ($Re$). Therefore, results can be generalised and used for various guide plates with different dimensions.
 \\
The experimental setup is sketched in figure \ref{fig:LB_1} for $E~=~1$.
In the setup, the air stream passes first a nozzle  before it is guided to the   guide plates. 
In front of these plates flow has a turbulence intensity of $TI~\approx~2.1\%$.
The vertical direction is not shown in figure \ref{fig:LB_standard}, since setup geometry does not change in this dimension.
\\
During the experiment, the number of inserted guide plates is varied, while holding $h_{in}~=~h_{out}$ fixed.
Therefore, the ratio $V$ can be expressed by  
\begin{eqnarray}
\label{eq:V}
V=\frac{W-n\cdot t}{(n+1)c},
\end{eqnarray}
with $W=6.7$~cm width of channel, $n~=~0,~1,~3,~6$ number of inserted guide plates, $t~=~1$~mm thickness of guide plates and chord length $c~=~\sqrt{2}\cdot r$ with radius $r~=~5$~cm.
Thus, small $V$ values are equivalent to a well guided flow in the corner.
Beside $V$ also the inflow velocity $u_\infty$ is changed to study $Re$-number dependency. 
The wake of the guide plates is measured by a single hot-wire probe.
\\
Results of these measurements are summarized in figure  \ref{fig:LB_standard}.
The figure  \ref{fig:LB_standard} a) shows a typical result of this investigation, the temporal mean velocity profile $\langle u(x)\rangle_t$ as well as turbulence intensity profile $TI(x)$ behind the guide plates.
As expected, the velocity modulations correspond to the plates positions. 
The velocity decreases in the direct wake of a plate and  turbulence intensity increases, however, slightly shifted.
\\
The efficiency of guide plates can be determined by holding ventilator power supply constant and change number of plates, while wake is recorded by hot-wire probe.
Preserving a constant ventilator power supply 
results in a {defined} and fixed driving pressure of the air stream, only the guiding features of the plates influence how much volume flux is present\footnote
{
The indicated velocity in the legend belongs to a calibration, where velocity out of the nozzle is related to control voltage of the ventilator. Thus, legend velocities are an orientation and not the real inflow velocities, which can be seen by the results in figure \ref{fig:LB_standard} b)
}.
The efficiency of guide plates can be expressed by the velocity profile and its average in space $\langle u\rangle_{(x,t)}~=~\langle\langle u\rangle_t\rangle_x$, which is related to the volume flux, 
since it is verified that flow features do not change significantly in the vertical direction. 
(For reasons of clarity subscripts are only written once.)
Figure \ref{fig:LB_standard} b) shows  $\langle u\rangle_{x,t}$ for different values of $V$ and for various inflow adjustments\footnote
{
An experiment performed with a fixed power supply of the ventilator is named an experiment with specific {inflow adjustment}
}.  
Two main statements can be drawn from these results in figure \ref{fig:LB_standard} b):
First, a change of $V$ results in a change of $\langle u\rangle$.
Second, the functional dependency of $\langle u\rangle$ and $V$ changes with the inflow adjustment and 
$Re$-number\footnote
{
Note, the investigation regarding the efficiency of guide plates with fixed inflow adjustment is only possible with special ventilators. Such a ventilator is characterized by a monotonous functional dependency between volume flux and counter-pressure, which means that in case counter-pressure increases volume flux must decreases. The used ventilators have such a dependency
}. 
\\
Figure  \ref{fig:LB_standard} c) shows the averaged turbulence intensity $\langle TI \rangle_x$ for various inflow adjustments and  $V$.
All $\langle TI \rangle$ developments decrease with decreasing  $V$. Thus, the more guide plates are installed the more reduces the level of wake turbulence. Note,  guide plates and the turning of the stream induce additional $TI$ in the flow, since inflow has $TI~\approx~2\%$.
Furthermore, the level $\langle TI \rangle$ decreases with increasing inflow velocity for almost all measurements.
\\
In a wind tunnel both quantities are of importance, thus a high ratio between $\langle u\rangle$ and $\langle TI \rangle$ shows that at the same time efficiency and flow quality are present.
The ratio is in units of velocity $[\frac{\rm{m}}{\rm{s}}]$. Therefore, one might interpret the ratio as some kind of {continuity measure of velocity}. 
The ratio $\frac{\langle u\rangle}{\langle TI \rangle}$ is shown in figure \ref{fig:LB_standard} d).
The four $\frac{\langle u\rangle}{\langle TI \rangle}$ developments show a dependency on $V$, which changes its functional form with changing inflow adjustment.
For inflow adjustment $u_\infty~\sim~3.5$~$\frac{\rm{m}}{\rm{s}}$ development is nearly constant, with a weak increase at lowest $V$.
Inflow adjustment \mbox{$u_\infty~\sim~6.7$~$\frac{\rm{m}}{\rm{s}}$} shows a development with decreasing $V$, which first increases and than stays almost constant.
Inflow adjustment $u_\infty~\sim~9.6$~$\frac{\rm{m}}{\rm{s}}$ shows an increasing 
ratio 
(more pronounced than for $u_\infty~\sim~6.7$~$\frac{\rm{m}}{\rm{s}}$) and stays than almost constant with decreasing $V$.
Inflow adjustment $u_\infty~\sim~12$~$\frac{\rm{m}}{\rm{s}}$ shows strict increasing $\frac{\langle u\rangle}{\langle TI \rangle}$ with decreasing $V$. 
Thus, the ratio still improves with decreasing $V$.
\\
In figure \ref{fig:LB_standard} d), one may see two regimes, in the one the ratio improves with decreasing $V$ and afterwards a regime is present where a decrease of $V$ does not lead to an improvement. 
At which $V$ the transition between the regimes happens and if these two regimes are present is a function of inflow adjustment and $Re$-number. 
For wind tunnel construction is this result of particular importance. 
In case one estimates $Re$-number of the guide plates, its possible to find a proper $V$ for a desired flow quality, which is a great benefit compared to the recommendation $V~\leq~0.25$ of \cite{Mehta}.   
\\
A further interesting extension of this investigation might be an experiment, which takes the inflow turbulence intensity into account as well.
Thereby, the question can be studied, if turbulence can be smoothed out with guide plates and to what extent.
Moreover, the investigated space phase can be scanned more largely and more in detail, thus the two regimes become more characterized in terms of $V$ and $Re$-number.
\\
Under the consideration of the reached results in combination with the features of the expanding corner (section \textit{g}),
the ratio of $V~=~0.29$ (with \mbox{$r~=~20$~cm}) is chosen for the circular shaped guide plates. 
The choice of this specific $V$-value is commented in detail in section \textit{g}.

\subsection*{g: Expanding corner}
\begin{figure}[h]
 \centering  
\includegraphics[width=0.6\textwidth]{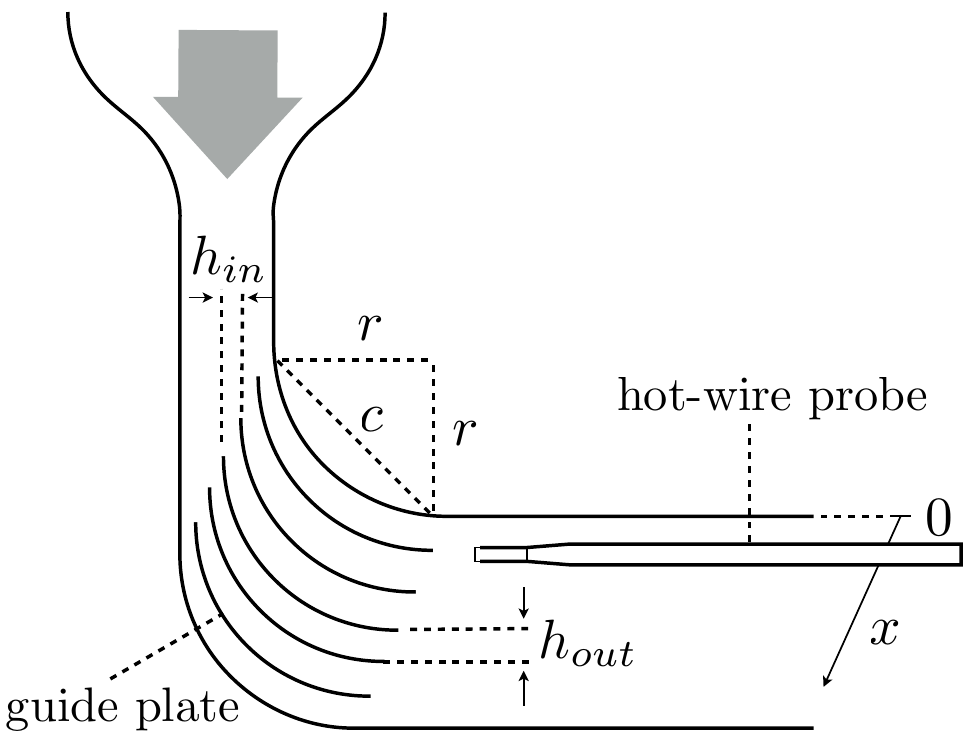}
\vspace{0.5cm}
\caption{Experimental setup for investigations on guide plates installed in an expanding corner with $E~=~\frac{h_{out}}{h_{in}}~=~2.3$. 
Massive walls are indicated by solide black lines. 
Grey arrow shows flow direction. 
Arrow at the hot-wire probe indicates the displacement x and its direction}
\label{fig:LB_2}
\end{figure}

\begin{figure}[h]
\centering  
a)\includegraphics[width=0.45\textwidth]{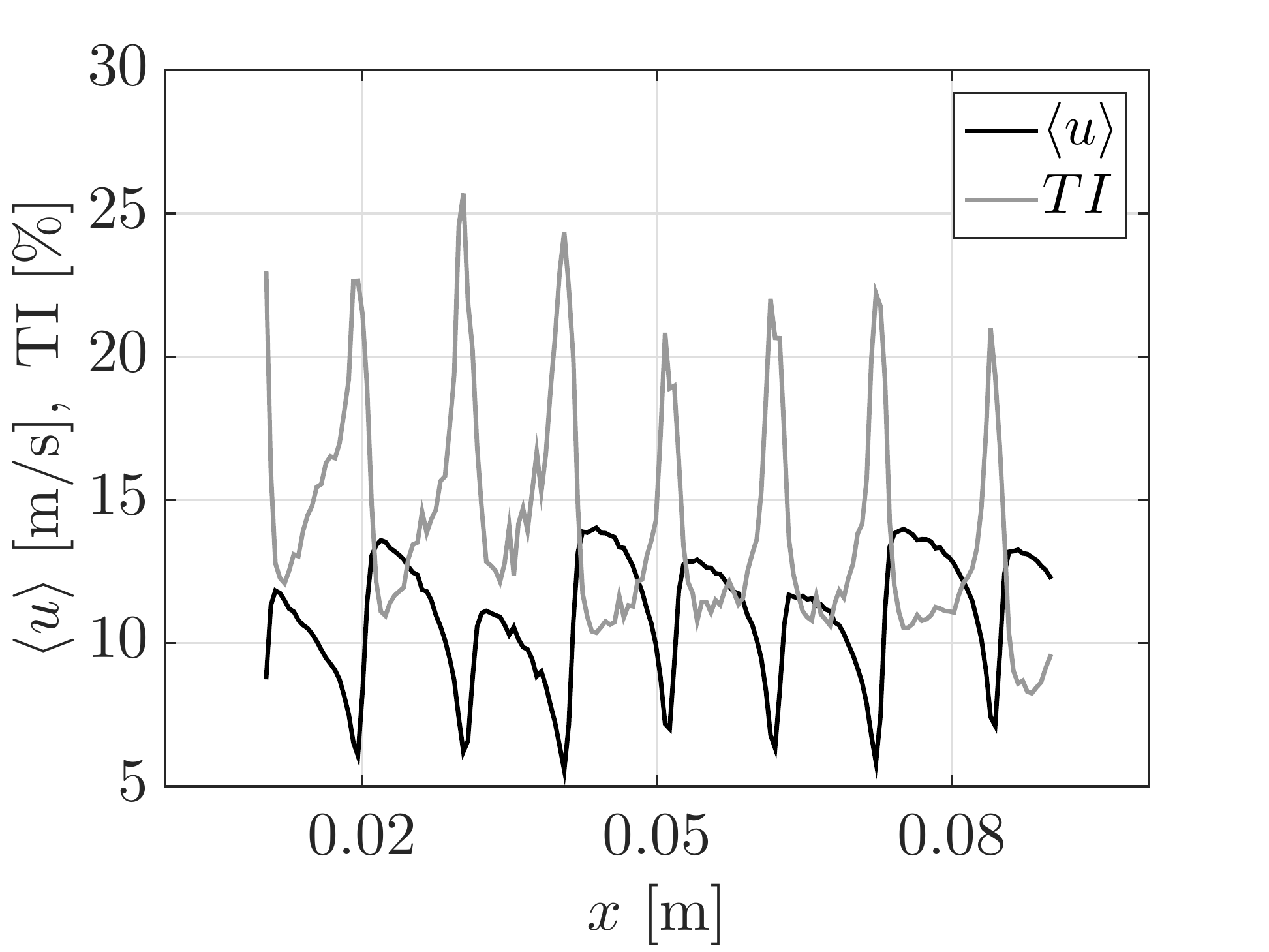}
b)\includegraphics[width=0.45\textwidth]{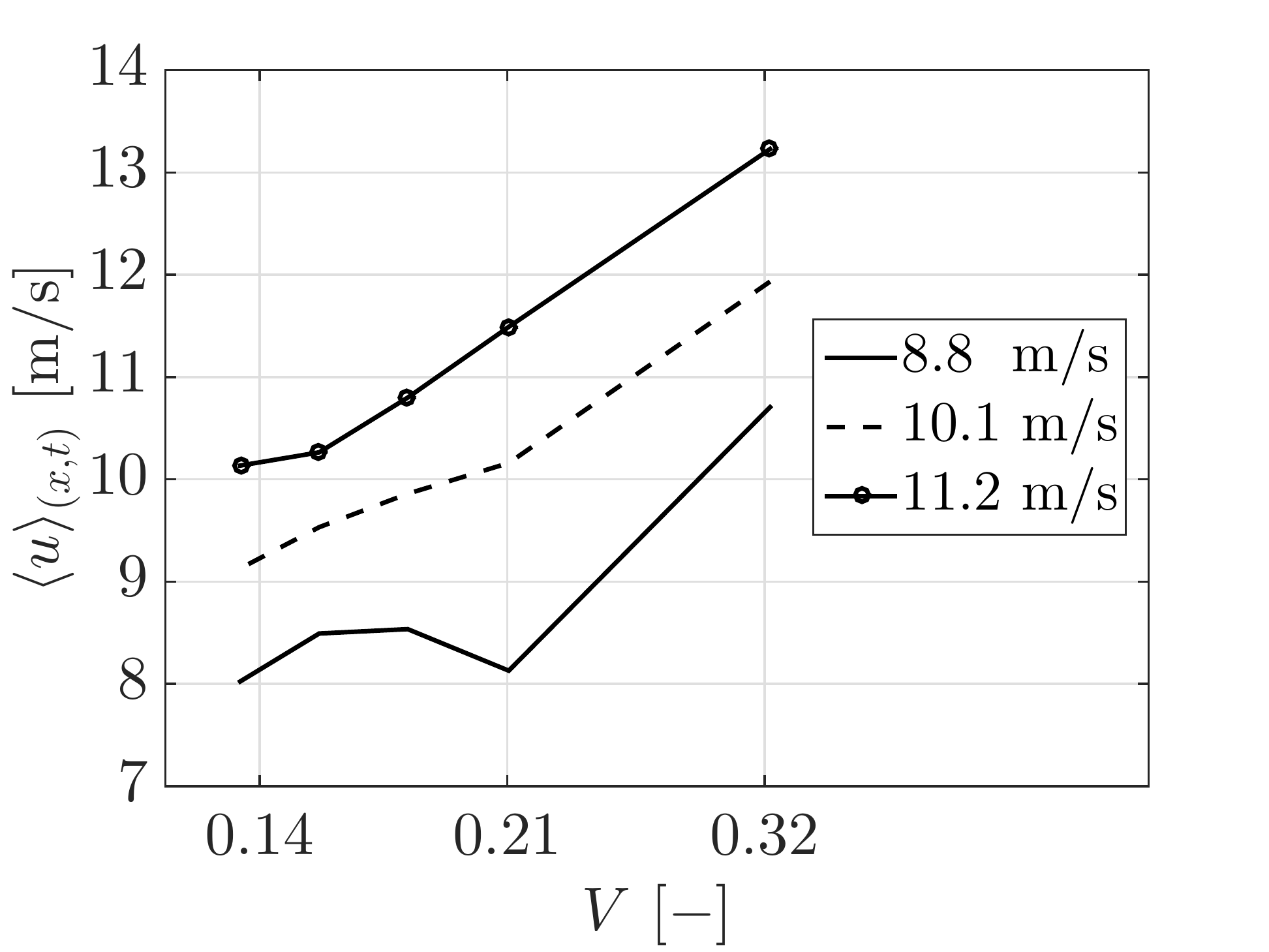}
c)\includegraphics[width=0.45\textwidth]{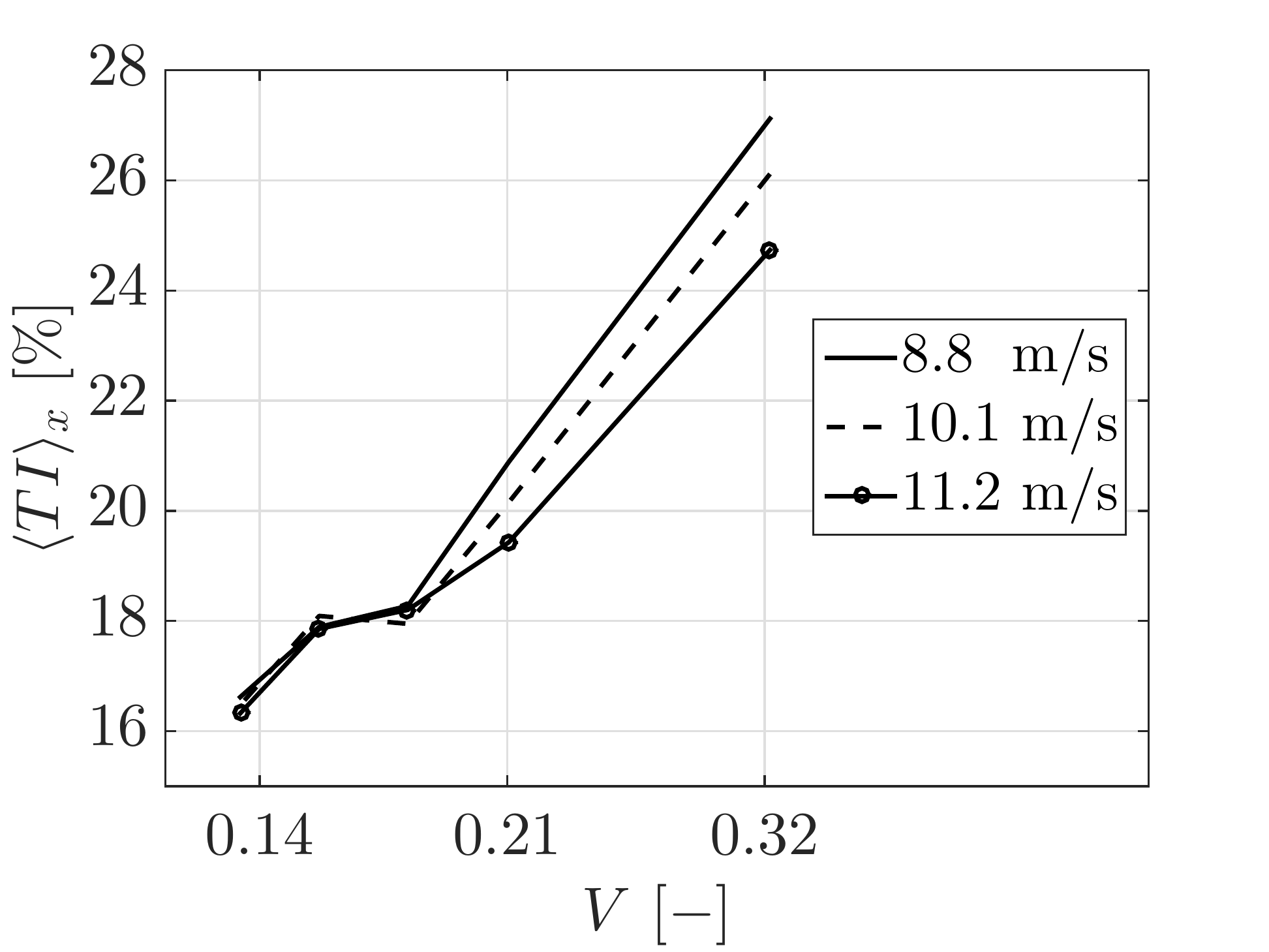}
d)\includegraphics[width=0.45\textwidth]{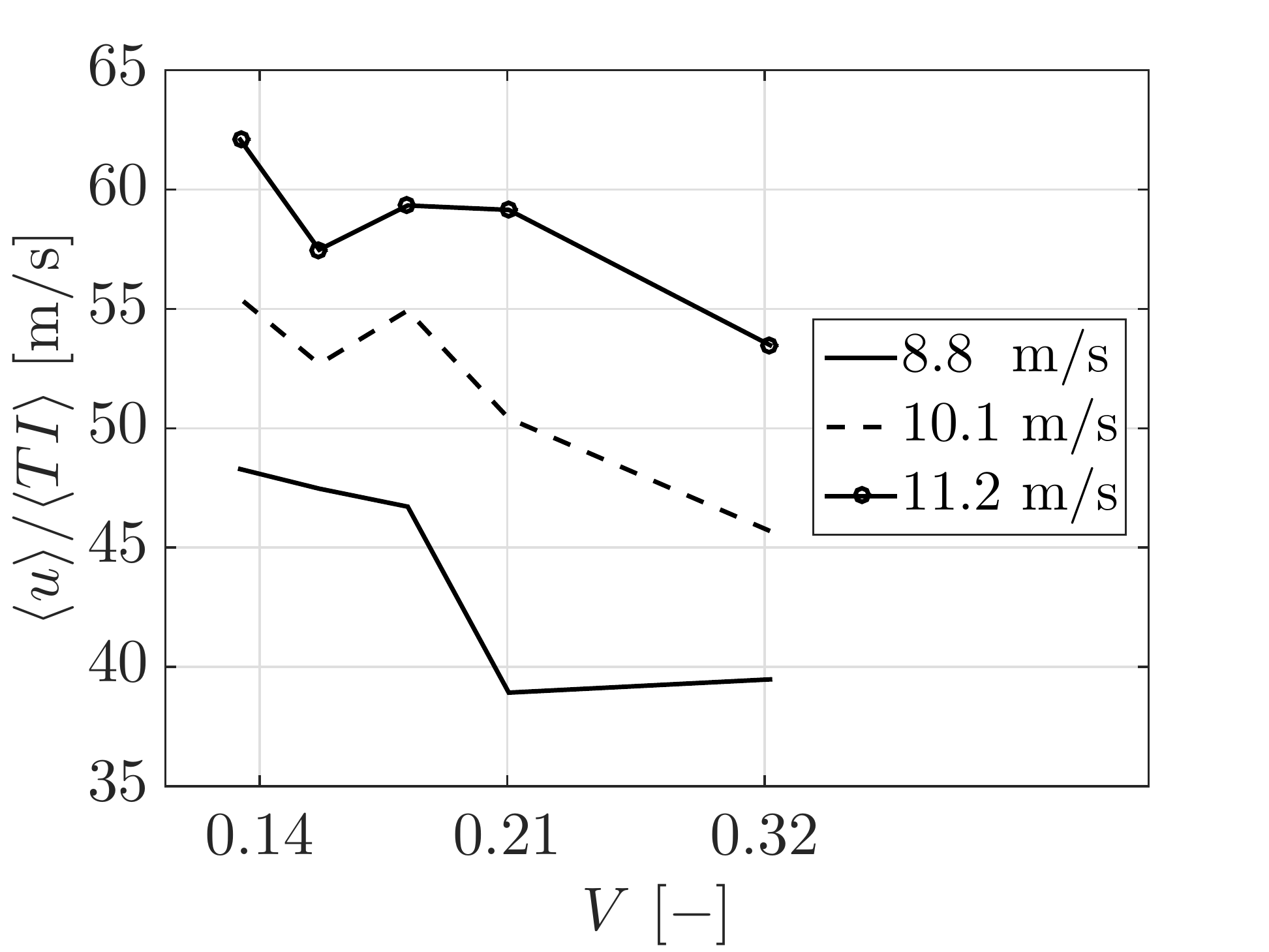}
\vspace{0.5cm} \caption{
Figures summarizes guide plates wake results with $E~=~2.3$.
a) temporal mean velocity profile  and turbulence intensity profile in the wake of guide plates, $V~\approx~ 0.14$,
b) spatial and temporal mean velocity with variation of $V$ and inflow adjustment
c) spatial mean turbulence intensity with variation of $V$ and inflow adjustment
d) ratio of spatial and temporal mean velocity b) with spatial mean turbulence intensity c)
}
\label{fig:LB_aufweitung}
\end{figure}

\begin{figure}[h]
\centering  
\vspace{1cm} 
\includegraphics[width=0.8\textwidth]{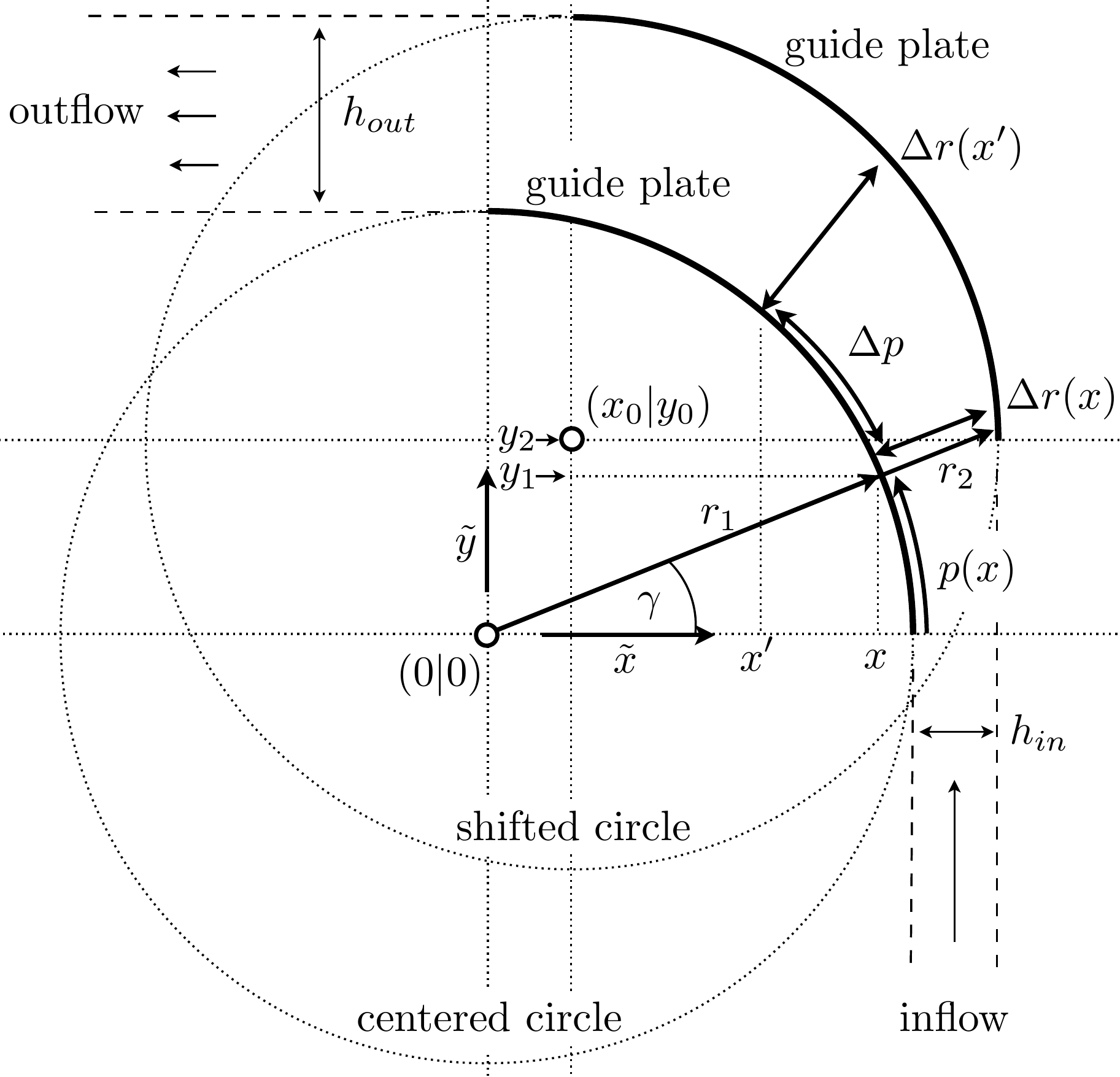}
\vspace{1.cm} \caption{
Drawing of guide plates and considered geometric quantities. Guide plates are turned around 180$^\circ$ in comparison to figure \ref{fig:LB_2}, which  facilitates the following considerations
}
\label{fig:LB_skizze}
\end{figure}

\begin{figure}[h]
\centering  
\includegraphics[width=0.6\textwidth]{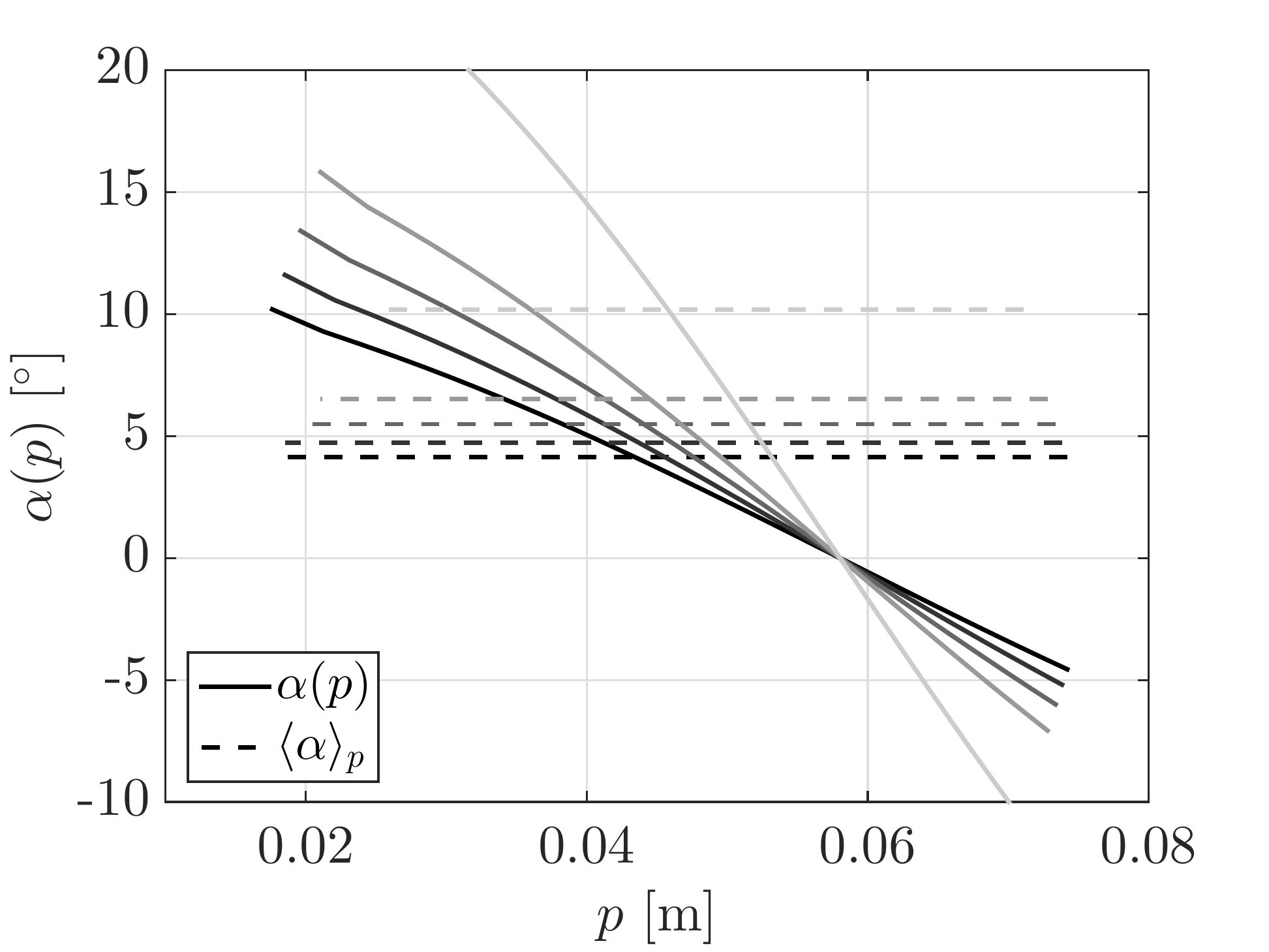}
\vspace{0.5cm} \caption{
Solid lines show opening angle $\alpha(p)$ of flow along its downstream path $p$ between expanding guide plates. 
Dashed lines indicate the mean opening angle $\langle\alpha\rangle_p$.The radius is selected as $r$~=~5 cm
}
\label{fig:LB_1_2}
\end{figure}

Wind tunnel component \textit{g} in figure \ref{fig:wk} is an expanding corner with the expansion $E~=~2.3$. 
Expanding corners are little discussed in literature, for instance by  Lindgren and Johansson \cite{Lindgren}, who report from corners with the expansion $E~=~1.3$. 
Lindgren and Johansson used guide plates with varying thickness, which is different to the here presented approach where plates are used with constant thickness, as mentioned above. 
Thus, an experiment is realized with an expanding corner with $E~=~2.3$. 
The experiment is performed in the same way as the experiment above with $E~=~1$.
Figure \ref{fig:LB_2} shows the modified experimental setup.
The distance ratio is set to $\frac{h_{out}}{h_{in}}~=~2.3$ during performing experiments.
The ratio is calculated by \mbox{equation (\ref{eq:V})} with the quantities $W~=~9.4$ cm width of expanded channel (downstream of guide plates), $n~=~3,~5,~6,~7$ and $8$ number of inserted plates, $t~=~1$ mm thickness of plates and chord length $c~=~\sqrt{2}\cdot r$ with radius $r~=~5$~cm. 
\\
The achieved results are summarized in figure \ref{fig:LB_aufweitung} likewise to figure \ref{fig:LB_standard}.
Figure \ref{fig:LB_aufweitung} a) shows a typically time averaged velocity and turbulence intensity profile in the wake of guide plates.
Velocity drops as well as turbulence intensity increases straight behind plates as expected.
The single velocity {jets} vary in height  and are non-symmetrical.
This shape of the wake velocity profile with its jets is different compared to wake velocity profile shown for a corner with $E~=~1$,  fig. \ref{fig:LB_standard} a).   
The most likely reason for this specific jet shape might be a flow separation between two guide plates.
\\
Figure \ref{fig:LB_aufweitung} b) shows  $\langle u\rangle_{(x,t)}$ for different values of $V$ and for various inflow adjustments.
Other inflow adjustments than in section \textit{f, l}, since the experimental setup was modified.
$\langle u\rangle_{(x,t)}$ results show a trend, if $V$ decreases $\langle u\rangle$ also decreases\footnote
{
Since more guide plates are installed relative to the channel width than in a constant turning corner, pressure drop is more significant and mean velocity decreases strongly with decreasing $V$
}. 
As expected, mean flow increases with increasing inflow adjustment.
Whether or not $\langle u\rangle$ as function of $V$  changes its functional  form in a systematically kind with changing inflow adjustment is quite uncertain.
\\
Figure  \ref{fig:LB_aufweitung} c) shows the averaged turbulence intensity $\langle TI \rangle_x$ for various inflow adjustments and  $V$.
All $\langle TI \rangle$ developments decrease with decreasing  $V$. 
Thus, the more guide plates are installed the more reduces the level of wake turbulence. 
For large  $V$, $\langle TI \rangle$ decreases with increasing $Re$-number.
For small $V$ is a universal regime present, where $Re$-number does not affect $\langle TI \rangle$ development.
Note, $\langle TI \rangle$ is much higher
for the expanding  corner than for the constant corner, discussed above.
\\
The ratio $\frac{\langle u\rangle}{\langle TI \rangle}$ is shown in figure \ref{fig:LB_standard} d).
The three $\frac{\langle u\rangle}{\langle TI \rangle}$ developments show 
a increasing ratio with decreasing $V$. The shape of the development is quite independent on the inflow adjustment, these are only shifted in height. 
Thus, turbulence intensity decreases faster than mean velocity. 
Therefore, a smaller $V$ is always beneficial in terms of turbulence intensity and for the \mbox{ratio $\frac{\langle u\rangle}{\langle TI \rangle}$. }
\\
The results show constantly changing wake features of guide plates as function of $V$.
Since the expanding corner is upstream of the settling chamber (component \textit{h}), the smallest $V$ is realized that could be installed, which is $V~\approx~0.162$ with \mbox{ $h_{in}$~=~2~cm} and $r~=~20$~cm. 
Guide plates with this $V$-ratio induce roughly 18\% turbulence intensity. 
Consequently, the previous constant corner (component \textit{f}) can induce a share of turbulence, which should be only below the level of the expanding corner as well as the guided volume flux should be maximized.  
With the wake results of figures \ref{fig:LB_standard} b) and c), $V~\approx~0.29$ is selected for the constant corner guide plates, with $r~=~20$~cm.
\\
\\
Furthermore, a few remarks are made on the experimental investigations and on the geometrical features of guide plates in expanding corners.
Measurement results in figure  \ref{fig:LB_standard} do not allow further statements on the dependency of $Re$-number.
Most likely, the experimental setup is not good enough 
for deeper interpretations,
since the velocity profile in figure \ref{fig:LB_aufweitung} a) shows already a strong variation between adjacent jets, although, guide plates are installed equally.
Thus, the wake profile seems to be very sensitive on slightly changing boundary conditions in expanding corners.
Consequently, a bigger and more precisely build up experiment may overcome such issues and more similar jets can be studied and further statements might get possible.
Beside a better experimental setup the investigations could be extended as proposed for the constant corner.
\\
For a deeper understanding how the flow gets expanded on its way between the guide plates, a local opening angle is determined.
Therefore, a few fundamental geometrical considerations are done below.
Finally, this local opening angle is compared to finding of diffusers, cf. \cite{Kuehle,Kline}.
Figure \ref{fig:LB_skizze} illustrates guide plates and considered quantities, which are discussed below.  
\\
The guide plates are characterized by their shape, which is circular. 
Two of these can be described by the Cartesian equation of a circle,
\begin{eqnarray}
\label{eq:Kreisform}
\label{eq:circle1}
r^2 &=& x^2 + y_1^2,\\
\label{eq:circle2}
r^2 &=& (x+x_0)^2 + (y_2+y_0)^2.
\end{eqnarray}
$x$ and $y$ are Cartesian coordinates, the displacement of the second circle corresponds to $x_0\sim1$ and $y_0\sim2.3$ and $r$ is the radius of circles.\footnote
{
Note, the Cartesian coordinate $x$ should not be confused with prior used displacement of the hot-wire probe, section \textit{f, l} and  \textit{g}
}
Equation (\ref{eq:circle1}) and (\ref{eq:circle2}) are functions for $y$, thus these different circles are distinguished by the subscripts, $y_1$ and $y_2$.
Transposing equation (\ref{eq:circle2}) to $y_2$, the distance $r_{2}$ of the shifted circle to the coordinate system origin can be expressed as function of $x$, 
\begin{eqnarray}
\label{eq:distance_1}
y_2(x) &=&  \sqrt{r^2 - (x-x_0)^2}-y_0,\\
\label{eq:distance_2}
r_{2}(x) &=& \sqrt{x^2 + y_2(x)^2}.
\end{eqnarray}
The distance of the center circle to the coordinate system origin is constantly $r_{1}~=~r$.
Thereby, the normal distance $\Delta r(x)$ between the circles can be expressed as
\begin{eqnarray}
\label{eq:Kreisform}
\Delta r(x) &=& r_{2}(x) - r,\\
 &=& \sqrt{x^2 + y_{2}(x)^2} -r\\
 &=& \sqrt{x^2 + \Big(\sqrt{r^2 - (x-x_0)^2}-y_0\Big)^2} -r.
\end{eqnarray}
The downstream path $p$ of the flow between the plates is expressed as
\begin{eqnarray}
\label{eq:Pfad}
p(x) &=& 2\pi\cdot r \cdot \frac{\gamma(x)}{360},\\
 \gamma(x)  &=&  \arctan \bigg(\frac{y_2(x)}{x}  \bigg).
\end{eqnarray}
$\gamma$ is the range of angles of a quarter circle, here applicable in the range of $11.8^\circ\leq\gamma\leq85.4^\circ$.
The local opening angel $\alpha(x)$
can be expressed by the normal distance $\Delta r(x)$ and the path $p(x)$
\begin{eqnarray}
\label{eq:op_ang_1}
\alpha(x)&=&  \arctan \left(  \frac{\Delta r(x') - \Delta r(x)}{\Delta p} \right)\\
\label{eq:op_ang_2}
&\hat{=}& \arctan \left(  \frac{\Delta r(x -\delta x) - \Delta r(x +\delta x)}{p(x-\delta x)-p(x+\delta x)} \right).
\end{eqnarray}
Equation (\ref{eq:op_ang_1}) is related to figure \ref{fig:LB_skizze}, whereas equation (\ref{eq:op_ang_2}) shows the {correct} infinitesimal form, where $\delta x$ is an infinitesimal displacement around $x$.
Figure \ref{fig:LB_1_2} shows the expansion of the flow along $p$ for the investigated $V$ values, according to eq. (\ref{eq:Pfad}) and (\ref{eq:op_ang_2}). 
Additionally, the mean opening angle $\langle\alpha\rangle_p~=~\textrm{atan}((h_{out}-h_{in})/p)$ is shown in figure \ref{fig:LB_1_2}. 
The grey scale corresponds from bright to dark to $n~=~3,~5,~6,~7$ as well as $8$ and  $V~\approx~0.32,~0.21,~0.18,~0.15$ as well as 0.14, and $\langle\alpha\rangle_p~\approx~10.2^\circ,~6.5^\circ,~5.5^\circ,~4.7^\circ$ as well as $4.1^\circ$.
The $\alpha(p)$ developments and  $\langle\alpha\rangle_p$  flatten with decreasing $V$.
It also shows that at first the flow undergoes a strong expansion. 
The strong expansion decreases with increasing path, and furthermore the expansion changes to an contraction at the end of expanding guide plates.
\\
The measurement results, fig. \ref{fig:LB_aufweitung}, show that flow features still change with decreasing $V$.
The geometrical consideration, fig. \ref{fig:LB_1_2}, suggests that flow separation takes place
at the beginning of guide plates, where a large opening angle is present.
The velocity profiles, fig. \ref{fig:LB_aufweitung} a), suggest that the flow does not attach again on the plates, since jets are asymmetric.
Thus, a substantial improvement might be shaped guide plates in an expanding corner. 
The shape should be in such a kind that the opening angle is constant along $p$ (like dashed lines in fig. \ref{fig:LB_1_2}), 
rather than a strong expansion at the beginning and moreover a contraction at the end of guide plates. 
In the case guide plates are installed with an constant oping angle one might compare their guiding features to the  results of linear diffusers, cf. \cite{Kuehle,Kline}.
The need of varying thickness guide plates is recognized in this study. 
But varying thickness guide plates are more difficult to build, therefore the simpler solution is selected with constant thick guide plates.
However, new experimental investigations or CFD simulations (Computational Fluid Dynamics) have to be performed to examine such a new concept and show their advantage.

\subsection*{h: Settling chamber}
The settling chamber of the wind tunnel is component  \textit{h} in figure \ref{fig:wk}. 
The dimensions of the settling chamber are $2.30\times2.88\times0.5$~ m$^3$ (height $\times$ width $\times$ downstream length).
As the name suggests, the flow settles down in the chamber, thus flow fluctuations (turbulence intensity) get damped. 
Therefore, two grids and one net are installed in the chamber, also named flow manipulators.
The damping process is caused by the quadratic pressure drop of the girds and the net ($\Delta p~\propto~u^2$).
Thus, a {relatively fast} gust is stronger slowed down than a {relatively slower} gust by these manipulators
and the flow field gets equalized.  
The quadratic dependency of the pressure drop on the velocity is also the reason why the settling chamber has typically the largest cross section of all wind tunnel components, since it results in the slowest velocity and the lowest pressure drop. 
Furthermore, large coherent flow structures (vortexes)
get partly broken by the manipulators and the more
filigree wake structures decay faster, cf. \cite{Reinke2016b}. 
\\
The  geometrical specifications of grids and net are selected according to the present flow features, cf. \cite{Atichat,Groth}.
Equally to the flow straightener, the ratio between integral length scale of the present turbulent flow and the mesh width is referred to be important.
In any case mesh width should be smaller than integral length scale (for the break down of vortexes).
Commonly, mesh width is selected five times smaller than integral length scale, cf. \cite[p.140, 146]{Groth1988} and \cite{Atichat,Mehta}, 
which is used for the selection of manipulators.
Another design parameter is the blockage of the manipulator. 
The manipulator blockage should be in a range between 30\%~-~40\%, \cite[p. 527]{Atichat}. 
\\
Moreover, also the $Re$-number should be considered by the manipulator selection. 
Therefore, Reynolds number is considered based on the bar width or the wire diameter.
Namely, the transition at $Re~\approx~40$ to $50$ should be avoided while changing flow velocity, 
since it corresponds to the laminar turbulent transition and a change of damping features of the manipulators, cf. \cite[p.511]{Atichat}.
Table \ref{table:data_Netz} summarizes  the characteristics of selected manipulators.
The selected manipulators fulfil the above considered geometrical relations to the incoming turbulent flow.
With regard to $Re$ no laminar turbulent transition is expected 
in the velocity range between \mbox{7.3~$\frac{\rm{m}}{\rm{s}}$}~$\le~u_\infty~\le~16.3$~$\frac{\rm{m}}{\rm{s}}$.\footnote
{
The specification of the velocity range is best practice in literature. 
Due to the fact that the settling chamber inflow is turbulent and velocity is partly lower or higher than the average velocity, the specified range is in reality smaller
}
\begin{table}[h]
\centering
\vspace{0.5cm}
\begin{tabular}{l|cccccc} 
  Name     & type                      & mesh                  & bar/wire        & spatial             & $u(Re~\le~40)$ & $u(Re~\ge~50)$ \\
                &                              &   width                 & diameter       &  blockage       &                          & \\
                &                             & [mm]                    & [mm]              & [\%]                  &   [$\frac{\rm{m}}{\rm{s}}$]   &  [$\frac{\rm{m}}{\rm{s}}$]            \\
  \hline
  grid 1   & regulare          &  30                   &  5           & 30.56              &       -            &1.5  \\
              & squared   &    &   &   &   &\\
    grid 2   & circulare               &    6                   &  1          & 37.01         &        -          & 7.3\\
              & holes    &    &   &   &   &\\
    net       & braided                  & 1.6                   &  0.36       & 39.94      &     16.3      &  -  \\
\end{tabular}
  \vspace{0.5cm} \caption{Summary of grids and net characteristics. Specified velocities $u$ correspond to test section inflow velocity, at manipulators velocity is $\approx$~8 times smaller (according to the nozzle contraction)}
\label{table:data_Netz}
\end{table}
\\
The optimal distance between manipulators is discussed in many works. 
For instance,  \cite[p. 517, 522]{Atichat} and \cite[p. 151]{Groth1988} suggest different distances between 2.5  to 10 (mesh width), to reach a proper relaxation of the turbulent flow.
Lindgren and Johansson \cite{Lindgren} report of much larger distances.
However, the suggestion by \cite[p. 522]{Atichat} is used where 3~-~10 mesh width are a proper distance and in particular a distance of 5 is recommended.
This ratio of 5 leads to a total settling chamber length of 0.3~m.
To keep the possibility of retrofitting, the settling chamber length is 0.5~m, thus additional manipulators can be installed.

\subsection*{i: Nozzle}
The nozzle of the wind tunnel is component \textit{i} in figure \ref{fig:wk}.
In the nozzle the flow gets accelerated as well as velocity fluctuations get damped.  
Both, the flow acceleration and the damping of fluctuations is mainly determined  by the contraction ratio ($CR$) of the nozzle.
The flow enters the nozzle over a cross section of  $2.30\times2.88$~m$^2$ and leaves the nozzle over a cross section of  $0.80\times1.00$~m$^2$, which results in a contraction ratio of  $CR~=~8.28$.        
Lindgren and Johansson \cite{Lindgren} show an estimation for the reduction of velocity fluctuations for a given contraction ratio.
Accordingly, longitudinal fluctuations are damped by $u'/66.7$, transversal fluctuations are damped by $v'/2.9$, the standard deviation is damped like $\sqrt{\langle u'^2 \rangle}/30.3$ and $\sqrt{\langle v'^2 \rangle}/3.3$, cf. \cite[pp. 21-22, eq. (11-14)]{Lindgren}.
Two effects are mainly responsible for the damping of fluctuations.
First, due to the velocity increase in downstream direction, longitudinal flow fluctuations get redistributed during the flow gets contracted by the nozzle.
Compared to the mean flow, fast fluctuations move faster downstream, up to a point where the fast fluctuation fits in the ambient mean flow. Vice versa, the same happens for slower fluctuations, which move relative to the mean flow backwards.
Second, the flow and its turbulent structures get stretched (also known as {vortex-stretching}), since flow gets continuously accelerated on its way downstream in the nozzle. Thereby, structures get thinner and more affected by viscosity, thus these structures decay faster. 
\\
The geometry of the nozzle's contraction is of substantial importance for the quality of the out coming flow.
Therefore, the suggestion by Lindgren and Johansson \cite{Lindgren} is used for a s-shape nozzle, see figure \ref{fig:wk}. 
Their considerations are adopted, since the present wind tunnel and their wind tunnel are similar in features (velocity range and tunnel dimensions).
Beside \cite{Lindgren}, further information to the nozzle and its features can be found in \cite{Mikhail,Downie,Borger}. These works are dedicated  to the shape and the dimensions of the nozzle as well as its optimization.
\\
The length of the nozzle mainly determines the uniformity of the velocity profile in the test section. A nozzle, which is too short, is characterized by a non-uniform velocity profile, which has relative high velocity close to the wall and less velocity in the center, thus the flow overshoots at the nozzle walls.
Investigations of Eckert and Kenneth \cite{Eckert1976} found  that a ratio between nozzle length $L_{nozzle}$ and nozzle inlet edge $B$ of 1 is a proper choice to reach a uniform velocity profile. 
Higher ratios show only small improvements.
Furthermore, 
\cite[p.476]{Mikhail} found that no flow separation happens up to a ratio of 0.5 inside this specific nozzle shape.
\\
Dimensions of the installed nozzle are $\frac{L_{nozzle}}{B_{broadness}}~=~\frac{2.3~\textrm{m}}{2.88~\textrm{m}}~\approx~0.8$ and $\frac{L_{nozzle}}{B_{height}}=\frac{2.3~\textrm{m}}{2.3~\textrm{m}}~=~1$.
Thus, we expect a weak non-uniform velocity profile but no flow separation.  
More information about flow features are shown in section  \ref{sec:Testsection}.

\subsection*{j: Test section}
\begin{figure}[h]
\centering
\includegraphics[width=0.8\textwidth]{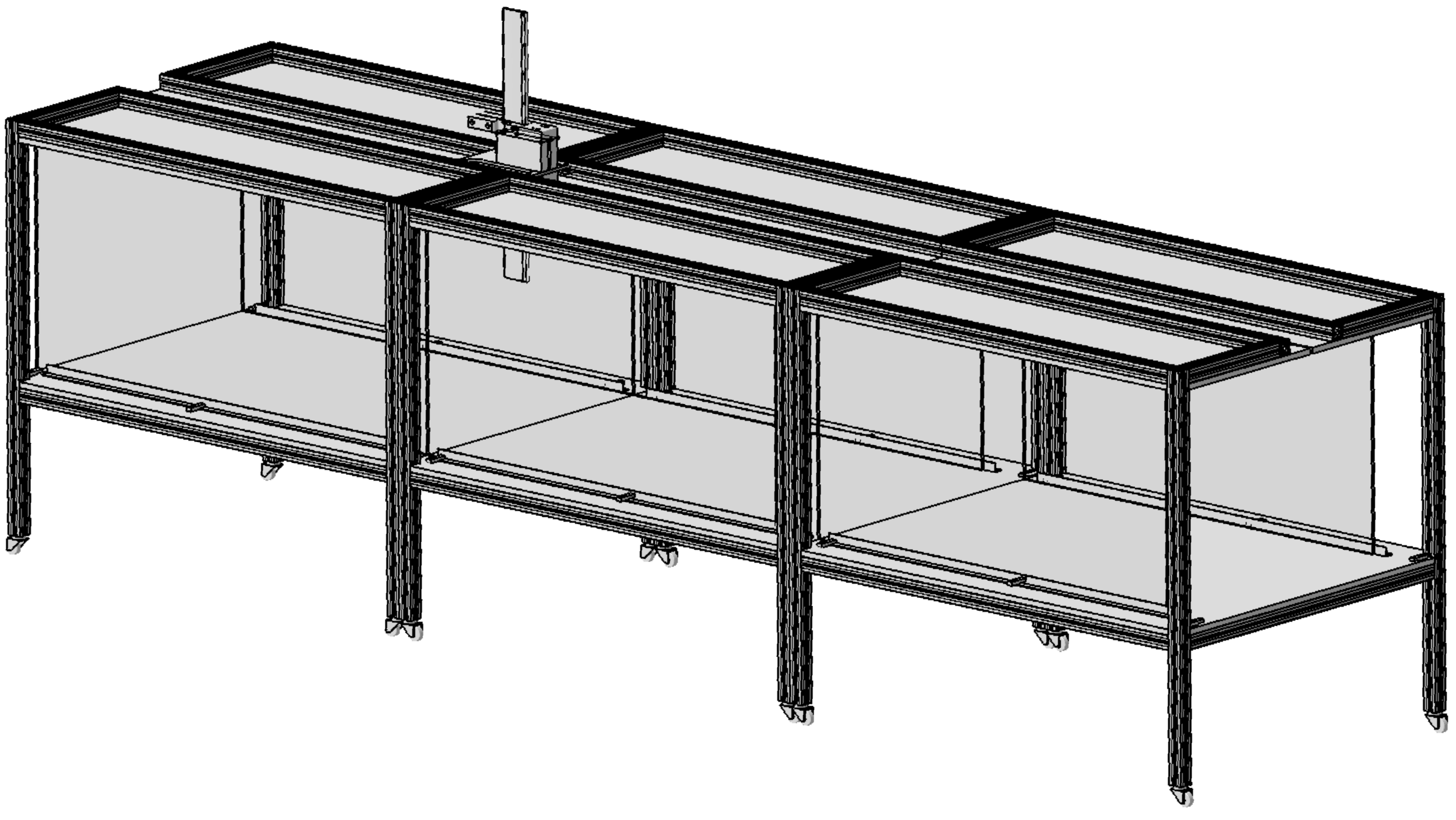}
\vspace{0.5cm} \caption{Technical drawing of the closed test section.  Due to a gap in the ceiling, a blade equipped with measuring probes (e.g. hot-wire probes)
can be inserted in the test section. The blade can be displaced by travel rails}
  \label{fig:geschlMessstr}
\end{figure}

The test section is component \textit{j} in figure \ref{fig:wk}. 
This component is designed for experimental investigation.
Two different test section setups are realizable and can be built up as required. 
First, an open configuration can be used with a length of 5.2~m, shown in figure \ref{fig:wk}.
In the open setup, the flow streams as a free jet from the nozzle to an inlet, from which the ventilator sucks in the flow.
Since the flow is a free jet, flow features significantly change on the way downstream, e.g. mean velocity and velocity profile as well as turbulence intensity.
\\
Second, a closed setup can be built up with dimensions of $0.80\times1.00\times 5.01$~m$^3$ (height $\times$ width $\times$ length), shown in figure \ref{fig:geschlMessstr}.
The closed test section is built up by three segments, each contains a measurement volume with dimensions of $0.80\times1.00\times1.67$~m$^3$ (height $\times$ broadness $\times$  length).
Therefore, the reconstruction is facilitated from an open setup to a closed setup and vice versa. 
Moreover, a shortened closed test section setup can be built up.
\\
In contrast to the open setup, flow features are much more constant in the closed test section. 
However, the natural boundary layer of the flow\footnote
{
Also known as viscous layer, since the flow is strongly effected in this layer by its viscosity. 
In this layer, the velocity changes from zero at the wall to the inflow velocity. 
In case a flow streams over a flat plate, the boundary layer is infinitesimal thin at the beginning of the plate and grows continuously downstream, e.g. \cite{Schlichting2000}
}
grows on its way downstream and makes it difficult to maintain flow features on a constant level within the test section. 
The growth of boundary layer is equivalent to an increasing velocity deficit at the wall.  
Since the volume flux is constant in a close test section
the flow outside of the boundary layer gets accelerated on its way downstream.
In some experiments, its necessary to conserve the central velocity, which is possible by slightly incline test section walls.
A typical inclination angle is $\alpha~\approx~\pm~0.2^\circ$, such adjustments and even more as well as negative inclination angles are possible with the constructed test section. 
Therefore, side walls can be moved and fixed in a certain position and a velocity increase can be balanced out.
One application of such adjustable walls might be an investigation concerning turbulence and its features (e.g. decay of turbulence) under different downstream velocity gradients ($\sim$ pressure gradients).
\\
Side walls are made of acrylic glass, therefore the test section is optically accessible (e.g. with laser-doppler anemometer).
Due to a gap in the entire ceiling, a blade equipped with measuring probes (e.g. with hot-wire probes) can be inserted in the test section.

\subsection*{k: Free jet collector and 3rd corner of guide plate}
The free jet collector and the 3rd corner of guide plates are component  \textit{k} in figure \ref{fig:wk}.
The cross section of the inlet is $2.25\times2$~m$^2$ (width $\times$ height). 
The cross section corresponds to a free jet expansion with an opening angle of roughly $\pm10^\circ$, cf. \cite{Schlichting2000}.
\\
Right behind the collector, a contraction corner is located (1:0.5), which turns the flow around $90^\circ$.
Installed guide plates have a distance between each other of \mbox{$h_{out}~=~8$~cm}, with a chord length of \mbox{$c~=~\sqrt{2} \cdot 20$~cm} is $V~\approx~0.28$. 
Since these guide plates contract the flow, the ratio of $V~\approx~0.28$ is considered as stable and uncritical. 
In particular, a lower turbulence intensity of the wake is expected in comparison to a constant turning corner.

\subsection*{m: Ventilator pre-nozzle}
The ventilator pre-nozzle is component \textit{m} in figure \ref{fig:wk}.
The contour of pre-nozzle  is shaped like a quarter circle with radius $r~=~0.625$~m. 
It contracts the flow in the horizontal direction by 2.25:1. 
Therefore, flow gets accelerated and equalized before it enters the ventilator.

\subsection*{n: Transition from squared to circular cross section}
\begin{figure}[h]
\centering  
\hspace{0.5cm}
\includegraphics[width=0.7\textwidth]{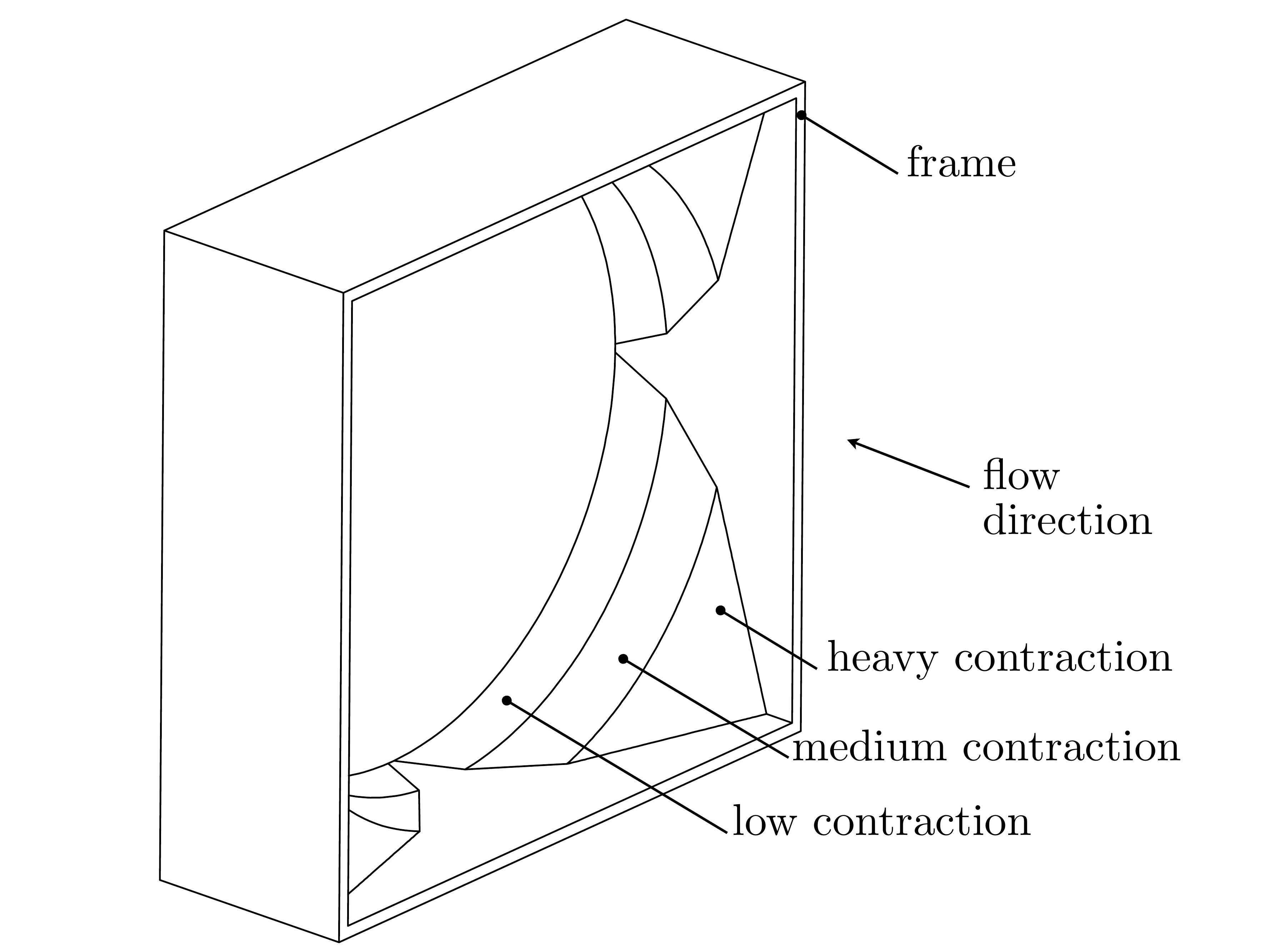}
\vspace{0.5cm} \caption{
Detailed view of component \textit{n}
}
\label{fig:component_n}
\end{figure}

The finial component is a transition from squared to circular cross section, which is \mbox{component \textit{n}} in figure \ref{fig:wk}. Figure \ref{fig:component_n} shows a detailed view of the component.
The contraction of the component is approximately 1:0.8, 
since the edge length of the squared inlet is 1.01~m as well as the diameter of circular outlet.
The transition is realized by four elements in the corners of the frame. 
These morph the squared to the circular cross section. 
Every element consists of three different transition and contractions steps.
Contractions are characterized first by a heavy contraction, second by a medium contraction and third by a low contraction. 
This transition from first to third contraction is inspired by a quarter circle, likewise to component \textit{m} the pre-nozzle.
This shape should minimize flow separation in front of the ventilator.

\section{Test section flow features}
\label{sec:Testsection}
\begin{figure}[h]
\centering  
a)\includegraphics[width=0.45\textwidth]{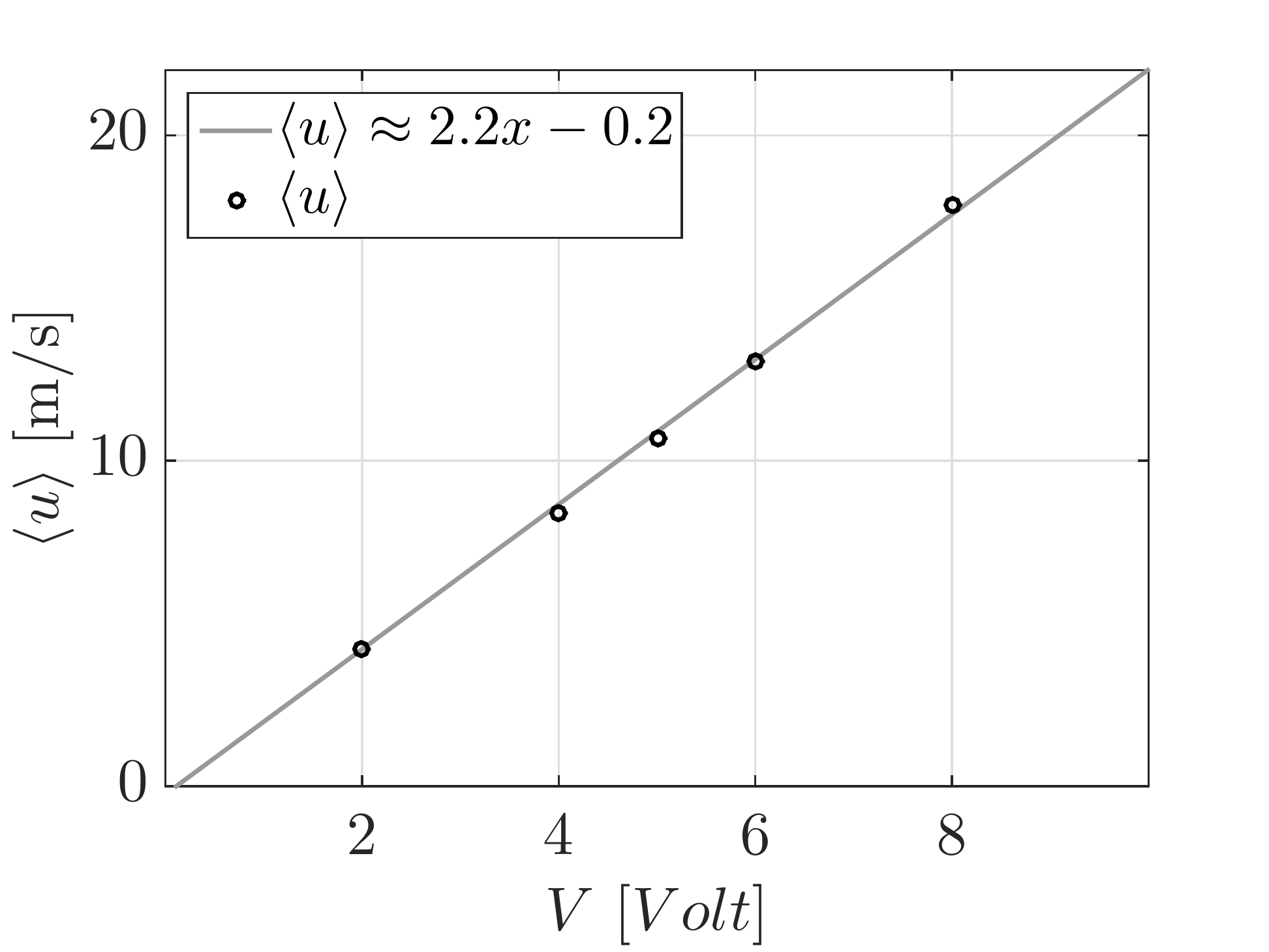}
b)\includegraphics[width=0.45\textwidth]{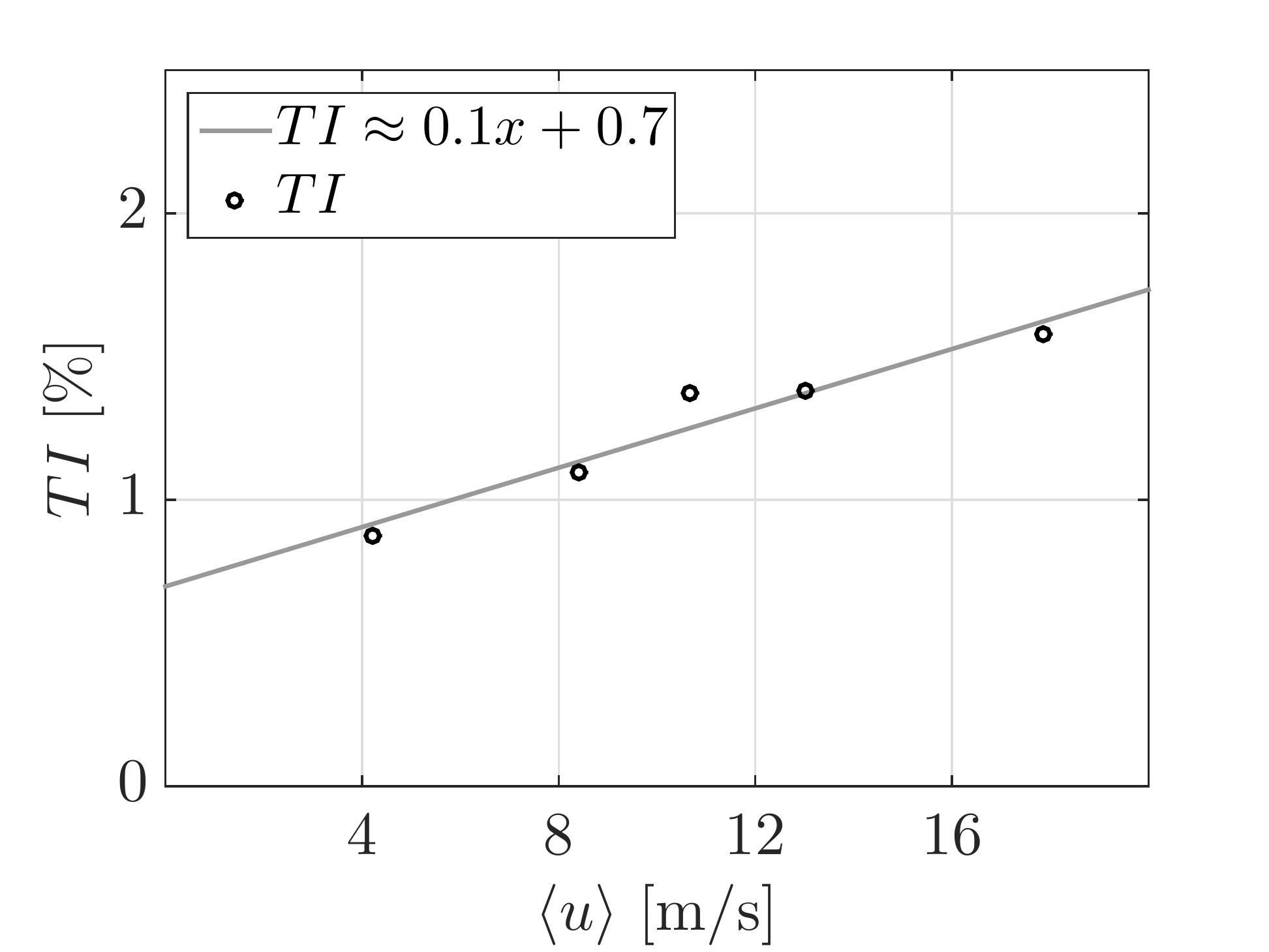}
c)\includegraphics[width=0.45\textwidth]{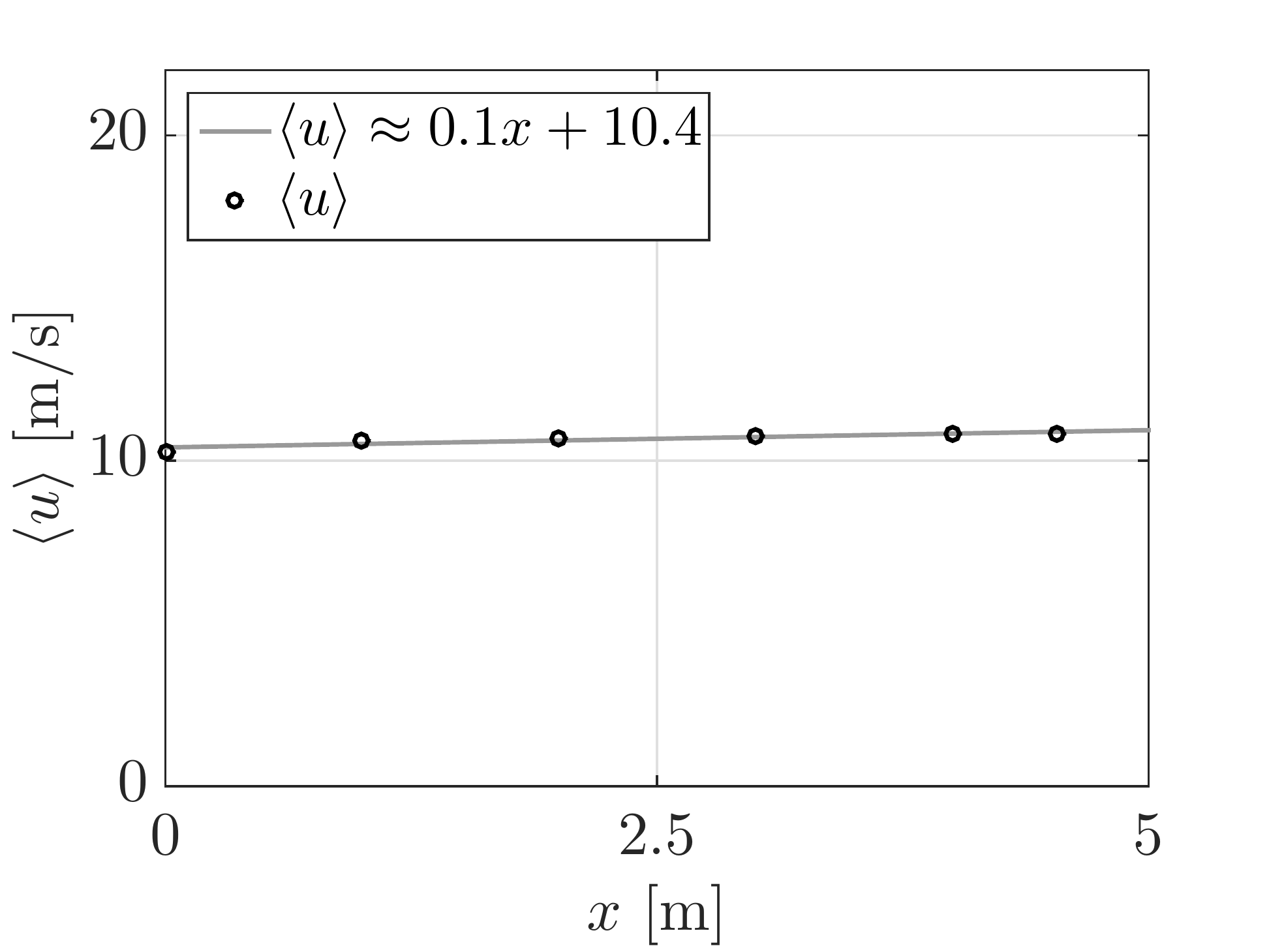}
d)\includegraphics[width=0.45\textwidth]{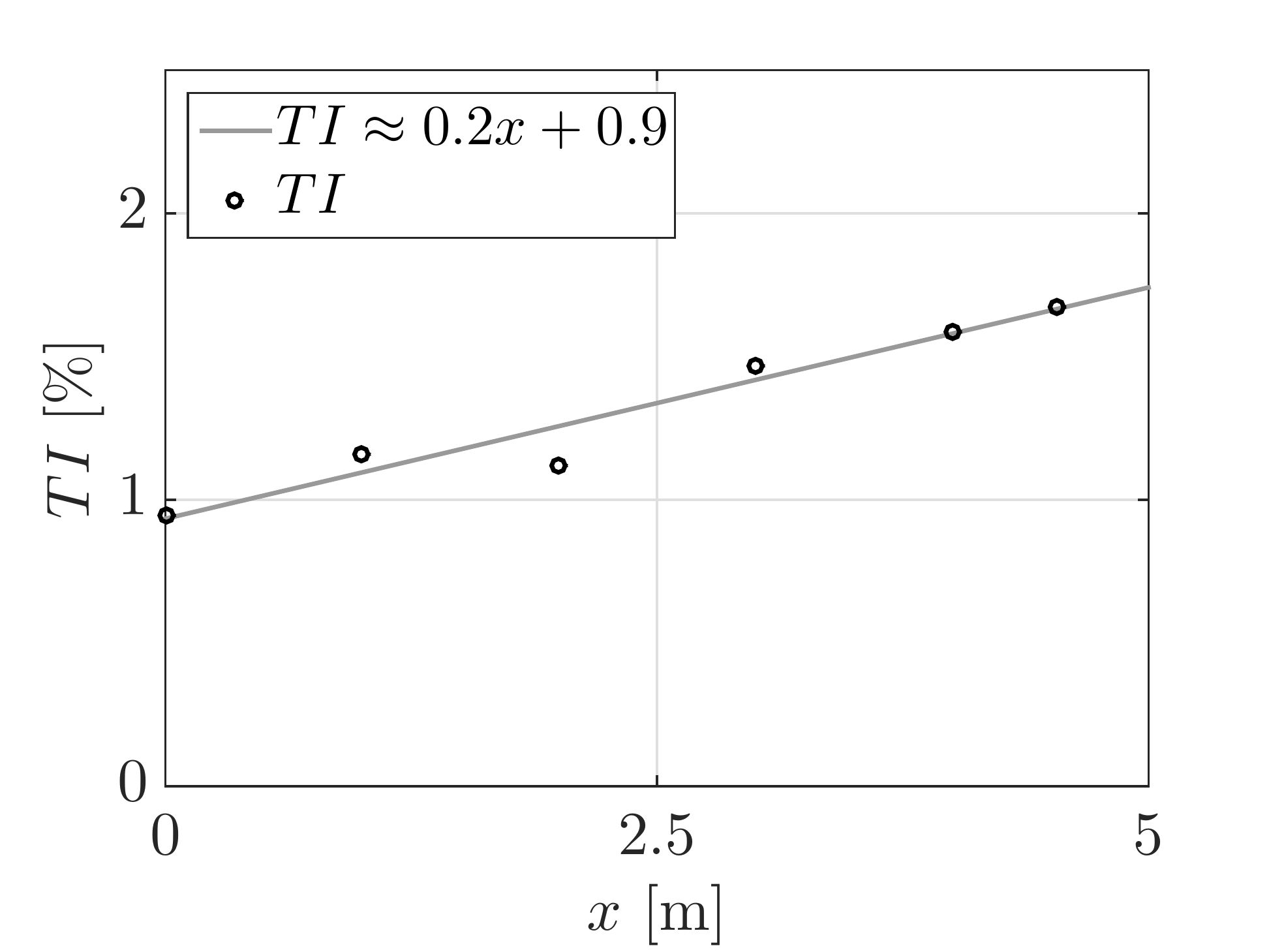}
\vspace{0.5cm} \caption{
a) temporal mean velocity as function of ventilators control voltage,
b) turbulence intensity over temporal mean velocity,
c) temporal mean velocity along downstream position,
d) turbulence intensity along downstream position.
a)-d) are measured at the beginning or along the centerline of the test section. 
Linear fit functions indicate dependencies between the shown quantities
}
\label{fig:WK_Kali}
\end{figure}

\begin{figure}[h]
\centering  
a)\includegraphics[width=0.45\textwidth]{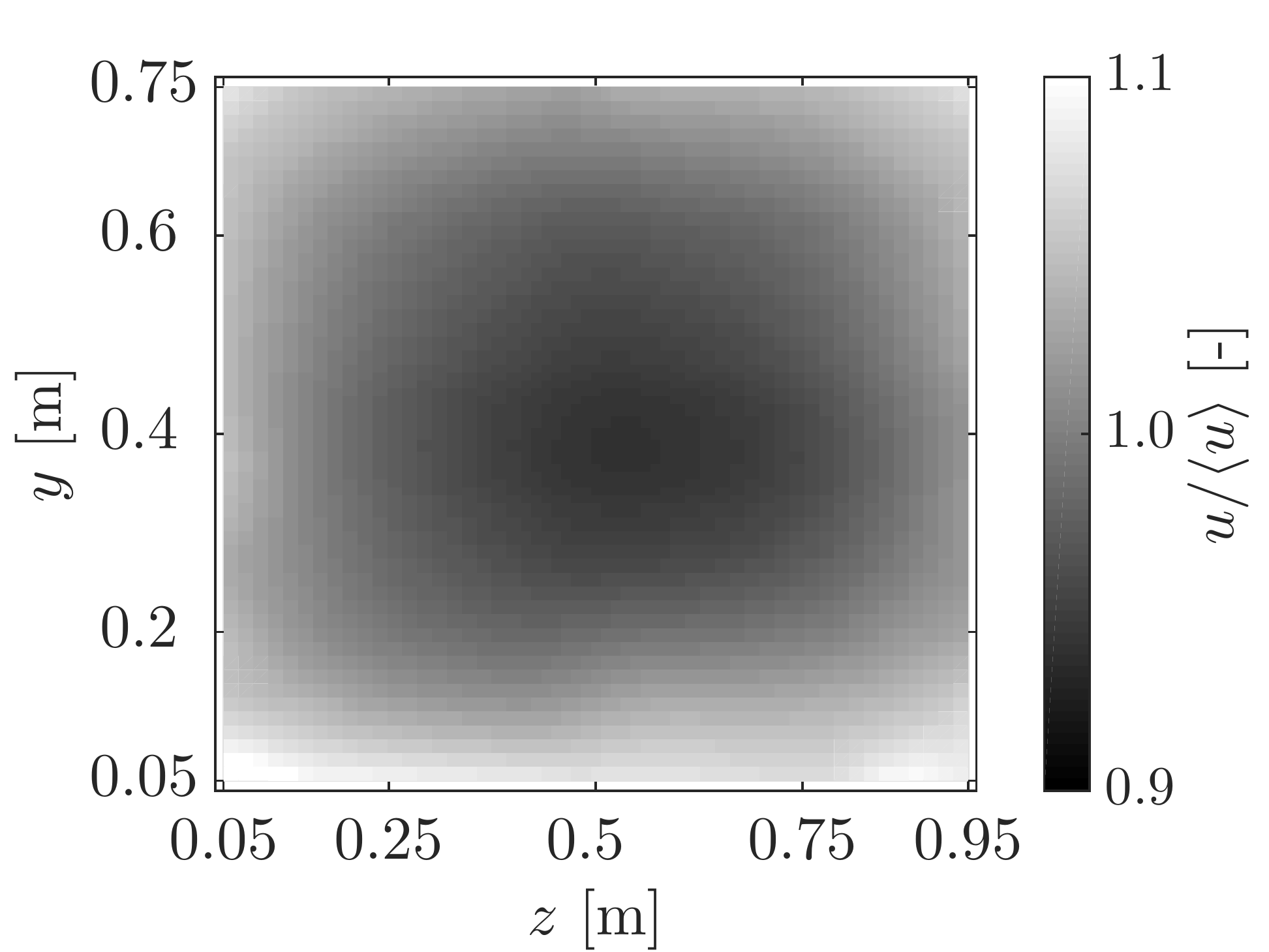}
b)\includegraphics[width=0.45\textwidth]{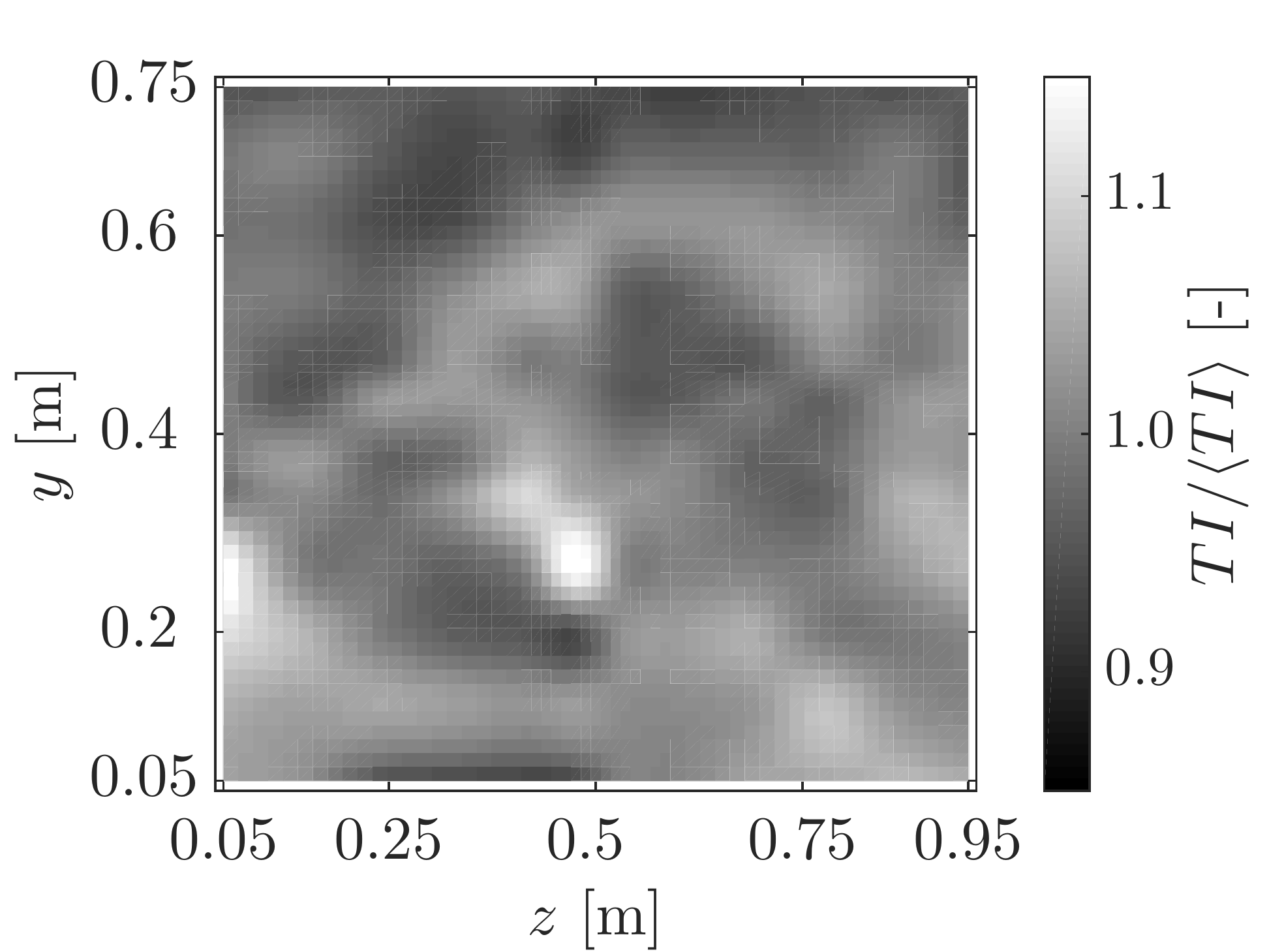}
\vspace{0.5cm} \caption{
a) Normalized velocity profile, b) normalized turbulence intensity profile. Both profiles are measured straight behind the nozzle outlet
}
\label{fig:U_TI_profil}
\end{figure}

Figure \ref{fig:WK_Kali} shows flow features in the closed test section. 
Measurements are performed with a single hot-wore probe.
Figure \ref{fig:WK_Kali} a) shows the temporal mean velocity as function of ventilator control voltage, measured at the beginning and at the center of the test section.  
Measurements agree with a linear fit function, $\langle u \rangle~\approx~2.2\cdot V - 0.2$.
The linear increase is expected, since control voltage is directly proportional to the rotation frequency of the ventilator rotor. 
However, this linear dependency and the maximal velocity of $\langle u_{max} \rangle~\approx~22$~$\frac{\rm{m}}{\rm{s}}$ indicate a low  overall pressure drop of the wind tunnel and a high efficiency.
A significant pressure drop leads to a non-linear dependency, since $\Delta p \sim u^2$ and causes also to a lower maximal velocity than the manufacturer specification of the ventilator, which is in fact even higher than its specification for $\Delta p~=~0$. 
\\
Figure \ref{fig:WK_Kali} b) shows turbulence intensity as function of the inflow mean velocity, measured at the beginning and at the center of the test section.
Turbulence intensity increases with increasing velocity approximately linearly,  $TI~\approx~$ \mbox{$0.1\cdot \langle u \rangle + 0.7$}.
\\
Figure \ref{fig:WK_Kali} c) shows how velocity changes downstream along the center line, with parallel adjusted side walls. 
This velocity increase, $\langle u \rangle~\approx~0.1\cdot x + 10.4$, might be related to the boundary growth, discussed above in section \textit{j}.
\\
Figure \ref{fig:WK_Kali} d) shows the turbulence intensity and its increase with downstream position, $TI~\approx~0.2\cdot x + 0.9$. 
Most likely, the increase can be explained with the growth of boundary layer thickness, 
which weakly influences also centerline measurements due to pressure fluctuations. 
Another reason for the increase might be related to the outlet of the closed test section, 
where the flow separates from the test section walls, which is also the origin for pressure fluctuations. 
\\
Figure \ref{fig:U_TI_profil} shows (a) the normalized velocity 
$u_{norm}=\frac{\langle u(t,y,z) \rangle_{t}}{\langle u(t,y,z) \rangle_{(t,y,z)}}$ 
and (b) 
normalized turbulence intensity profile $TI_{norm}=\frac{TI(y,z)}{\langle TI(y,z) \rangle_{(y,z)}}$.\footnote
{
Velocity measurements were conducted at $12\times12$
 evenly distributed positions over the nozzle outlet. 
In postprocessing, the measurements are interpolated along the cross section
}
Both profiles are exemplary for various velocities.
At first sight, both are quite uniform.
The uniformity of these profiles can be quantitatively expressed in their standard deviation, which is $\sigma(u_{norm})\cdot100~\approx~4\%$ and $\sigma(TI_{norm})\cdot100~\approx~4\%$.\footnote
{
The measurement is performed with four single hot-wire-probes. 
Each of which measured one quadrate of the profile, which might be the reason why the four quadrates slightly contrast in figure  \ref{fig:U_TI_profil} b).
}
Thus, also from a quantitative point of view are both profiles quite uniform.
However, the velocity profile overshoots close to the walls even though recommendation concerning the nozzle length is respected, at least in horizontal direction. 

\section{Summary and conclusions}
Here, the design and construction of a wind tunnel is described.
Where in particular, the test section length is maximized in a laboratory with {limited} dimensions,
which results in an unusual wind tunnel design.
The requirement for this unusual design are acceptable turbulence intensity and a uniform velocity profile as well as a proper maximal flow velocity in the test section for turbulence experiments. 
The design is characterized by two large flow expansions, namely a wide angle diffuser and a strong expanding corner. 
\\
It is referred in literature that such a wide angle diffuser leads to unwanted flow separation. 
To prevent flow separation and to keep the flow attached at the walls of the diffuser, it is separated by additional walls, thus the main diffuser is divided in four sub-diffusers.
It turns out that these additional walls are highly advantageous in adjusting the wake velocity profile of the diffuser.
Thereby, the non-homogenous inflow of the diffuser becomes equalized within the diffuser and its wake gets homogenous. 
\\
While designing the tunnel, it becomes obvious that the existing knowledge regarding guide plates in turning corners is rather limited and in particular the knowledge about the installation of guide plates in expanding corner.
Consequently, two experimental investigations were performed.
Here the turning corner wake is analysed with regard to velocity and turbulence intensity as well as by a new quantity the ratio of both\footnote
{
The ratio indicates how efficient the flow is guided by the plates, 
since a high ratio is related to a high volume flux with relatively low turbulence intensity
and it is suggested that this ratio is a key parameter in wind tunnel construction
}, 
$\frac{\langle u \rangle }{\langle TI \rangle }$.
In a first investigation a regular constant turning corner is studied.
Flow characteristics show a strong dependency on the relative distance of adjacent guide plates $V$ as well as on the flow velocity.
In particular, $\frac{\langle u \rangle }{\langle TI \rangle }$ as function of $V$ shows a systematic dependence, which is characterized by two regimes. 
In the one regime a closer distance of guide plates is beneficial for the ratio and 
after a specific distance a second region is present, 
where the ratio does not increase, 
thus a closer distance does not improve $\frac{\langle u \rangle }{\langle TI \rangle }$.
The relative distance $V$ where this transition happens is a function of velocity and $Re$-number. 
This transition might be related to attached and separated flow states, i.e. in case flow is attached between the plates the ratio starts to saturate.
In contrast to the literature recommendation ($V\leq0.25$), these results ($\langle u \rangle$,  $TI$ and  $\frac{\langle u \rangle }{\langle TI \rangle }$) form the basis for a more valid designing of turning corners.
\\
A second investigation is dedicated to a strong expanding corner.
As expected, the experiments show quite different wake features.
Variation of $V$ varies strongly mean velocity and turbulence intensity of the wake of the guide plates. 
Interestingly, turbulence intensity becomes a {universal} feature, 
where only $V$ affects turbulence intensity level, 
but not the inflow velocity. 
The turbulence intensity decreases stronger than velocity with decreasing $V$.
Thus, the ratio $\frac{\langle u \rangle }{\langle TI \rangle }$ increases with decreasing $V$. 
In contrast to results of the constant corner, the ratio shows only one increasing and improving regime.
Geometrical considerations show the local opening angle of the flow while crossing guide plates.
This angle is above $10^\circ$ at the beginning of plates, even for the smallest $V$,
thus flow separation is expected, cf. \cite{Mehta}. 
This assumption of flow separation gets confirmed by the asymmetric wake velocity profiles.
Furthermore, it may indicate that variable thick guide plates are beneficial in expanding corner.   
This new findings about wake features of guide plates and its geometrical features in expanding corner enable an improved design of such turning corners.
\\
Wind tunnel characteristics are presented in terms of velocity and turbulent intensity features within the closed test section.
Velocity increases linearly with control voltage and rotor rotational frequency. The maximal velocity is \mbox{22~$\frac{\rm{m}}{\rm{s}}$}, which is above the manufacturer specification. 
Therefrom, one can conclude that the overall pressure drop of the wind tunnel is very low. 
Turbulence intensity increases in a linear fashion with increasing velocity, which ranges between  0.7\%~$\le~TI~\le$~2.9\%.
All in all these characteristics conform with predefined features, which are required for the planed turbulence experiments.
\\
However, the wind tunnel construction still leaves room for improvements, 
in terms of maximal velocity and minimal turbulence intensity.
Since the design of ventilator rotor blades and their surface roughness is not optimal, new blades might improve volume flux as well as uniformity of outflow velocity profile of the ventilator.
Additionally, the angle of attack of the blades might be readjusted, to improve the operating point of the ventilator in this wind tunnel setup.
Therefore, a higher maximal velocity might be reached and a certain velocity with a lower rotational frequency of the rotor (less noise induction). 
\\
The level of turbulence intensity can be significantly reduced with additional components at six locations in the tunnel.
First, as mentioned, an improved rotor blade design.
Thereby, large velocity gradients might be reduced and 
a more uniform flow with less turbulence is present right behind the ventilator.  
The second location is the wide angle diffuser. 
Additional walls in the diffuser (vertical as well as horizontal) decrease the opening angle and have a similar functionality as honey combs.
Third, the diffusers wake can be further equalized by a grid. 
Furthermore, such a grid may lead to an improved performance of the following honey combs, cf. \cite{Mehta}.  
Fourth, redesign of the expanding corner with new variable thick guide plates. 
The fifth location is the settling chamber, where more grids and/or nets can be installed.  
Sixth, a grid or net can be installed at the outlet of the closed test section.
\\
Concluding, this unconventional wind tunnel design shows a solid overall performance, which becomes clear 
in velocity and turbulence intensity features.
However, since the design is not completely exploited, this kind of design may reach equivalent flow features as wind tunnels of conventional design. 
Thus, the here presented work is a starting point for such unconventional wind tunnels, which allows saving and optimizing available construction space.

\section*{Acknowledgments}
I like to acknowledge Agnieszka H\"olling and J\"org  Dapperheld for fruitful discussions and technical support.
Furthermore, I want to thank
Heike Oetting (measurements in sec. \ref{sec:Testsection}),
Lars Varchmin (measurements in sec. \ref{sec:components} \textit{f, g} and \textit{l}) and
Thibaut Piquardt (measurements in sec. \ref{sec:components} \textit{d})
 for provided measurement data and fruitful discussions.
Moreover, I want to thank all colleagues who assisted  in building up the wind tunnel and for all the fruitful discussions with
Andr\'e Fuchs,
Christina He\ss eling,
Daniel Strutz,
Dominik Traphan,
Gerd G\"ulker, 
Gerrit Kampers,
Hendrik Hei\ss elmann,
Ingrid Neunaber,
Jan Frederick Unnewehr,
Jannik Schottler,
Joachim Peinke,
Jarek Puczylowski,
Jean-Daniel R\"uedi,
Lars Kr\"oger,
Michael H\"olling,
Natascha Hedrich,
Stanislav Rockel and
Tim Homeyer.
Finally, the workshop staff of the University of Oldenburg, who help beyond expectations.
Without all of this help the built of the turbulent wind tunnel Oldenburg would not have been possible.

\bibliographystyle{plainnat}
\bibliography{WT_arxiv_nur_pdfs}

\end{document}